# Reliable Uncertainty Estimation for Machine-Learned Multigroup Cross Sections Without Retraining

Joshua Nichols[a], Valerio Mascolino[a], Andrew Osborne[a,*]

[a]The Colorado School of Mines, 1012 14th St., Golden CO, USA

*Corresponding Author: osbornea@mines.edu

**Abstract**

Uncertainty quantification in machine learning models is essential for nuclear energy applications. While machine learning models can make both uncertainty and point predictions, these models are often significantly more difficult to train jointly than models that make point predictions alone. Split conformal prediction is a family of methods for computing distribution-free prediction intervals by calculating nonconformity scores corresponding to a target quantile in a calibration dataset. Although conformal prediction only guarantees marginal coverage, normalizing nonconformity scores enables predictive interval widths to be calculated on a per-sample basis, providing an approximation to conditional coverage. In this work, we connect uncertainty predicting neural networks to pre-trained networks that estimate shielding factors in multigroup neutron cross sections. We train the uncertainty models on the same OpenMC data used by the pre-trained models, whose weights and biases remain fixed. The outputs of the uncertainty models are the standard deviations of the residual, separated into contributions from the model and tally uncertainty. Their corresponding variances are combined in quadrature to provide a normalizing factor for conformal prediction, providing interval widths that are adaptive on a per-sample basis. Baselines for comparison are established using Mondrian conformal prediction, and by using the ground truth OpenMC uncertainties as normalizers in the conformal framework. We use predicted uncertainty interval widths and the severity of predictions that overshoot the interval widths as bases for comparing our normalizer with the baselines. We find that our method dramatically reduces both interval widths and the degree to which predictions fall outside these widths.

## 1. Introduction

Uncertainty quantification (UQ) in modeling and simulation (M&S) of nuclear reactors and fuel cycles is well established with a long history and comprehensive literature. While machine learning (ML) has shown significant potential and utility in nuclear engineering (NE) applications including M&S, UQ for ML in this context has received relatively less attention. To the authors' knowledge the first investigation into UQ in ML for NE specifically was published in 2005 [1]. This article described analytical methods for estimating prediction intervals applied to models including artificial neural networks (ANNs) in online monitoring of nuclear power plants. Although related work in this topic has been published sporadically since 2005, in recent years following the mainstream adoption of ML the NE community has increasingly recognized its importance. A recent article [2] highlighting the importance of UQ in ML for NE includes a comprehensive review of work done in this topic. The article also shows that this topic has received recent attention from national and international institutions such as the US Nuclear Regulatory Commission (NRC) and the Organisation for Economic Cooperation and Development's (OECD) Nuclear Energy Agency (NEA).

The complexity of the internal mechanisms of many popular ML models such as ANNs prevents systematic analysis from achieving complete and certain characterization of their accuracy and limits in a straightforward way. The development of UQ methods has nonetheless been pursued to provide estimates of prediction confidence for black-box ML models. In published NE studies specifically, methods include Monte Carlo dropout [2], [3], [4], [5], [6], Bayesian neural networks [3], [6], [7],[8] and Gaussian processes [4], [7], [9], [10]. A disadvantage of these UQ methods is that they require incorporation into the ML model architecture before training and optimization. Deep ensembles [6], [7], [11], [12], mean-variance estimation [12], and quantile regression [13] have also been successfully applied in NE. These uncertainty aware ML models can be coupled to a trained, optimized mean predicting model in a subsequent step. Optimizing ML models requires significant manual effort and expertise due to the large design space of ANN topologies and training hyperparameters. Consequently, the advantage of post-hoc UQ methods is that training and optimization for the mean-predicting component does not need to be repeated. Additionally, it has been shown that training ML models to make mean and variance predictions simultaneously is more difficult than training a variance head in a subsequent step [14], [15]. The benefits of post-hoc UQ methods, however, are not the focus of the cited studies.

Conformal prediction (CP) is another family of UQ methods that can be coupled to a previously trained model in general. Split conformal prediction (SCP) enables prediction confidence intervals to be calculated in unseen data samples [16], [17]. Importantly, the prediction intervals are guaranteed [18] to cover a specific fraction of unseen samples under the assumption of exchangeability. This assumption requires that unseen samples for which predictions and confidence intervals are sought are drawn from the same underlying distribution that generated a calibration data set. Interest in CP methods is emerging in NE, with recent work including UQ in void fraction [8] and swelling [19] predictions. A limitation of CP, however, is that in general it cannot guarantee coverage of prediction intervals conditioned on individual samples [20]. Instead, the guarantee provided by CP is restricted to marginal coverage, where prediction intervals correspond to the specified level of coverage only on average [21]. Mondrian conformal prediction (MCP) extends SCP by stratifying the calibration data along a chosen axis, guaranteeing coverage in each stratum [21], [22]. The limitation of marginal coverage in these CP methods can often be mitigated, however, using normalization to allow prediction intervals to vary in width according to an estimate of a prediction's difficulty [22]. A recent study [23] employs normalized CP in the context of operator networks for predicting thermal-hydraulic quantities in nuclear reactors. The authors use the standard deviation computed by a dropout-ensemble of operator networks as a basis for a normalizer to enable adaptive prediction intervals. The normalizer is related to the spread in predictions observed by sampling over multiple inferences, adjusting network weights in each inference using a Monte Carlo approach.

Building on work in [23], here we develop a normalizer for CP with per-sample adaptivity enabled by two independently trained neural networks that predict the model (epistemic) and data (aleatory) uncertainties. These uncertainty neural networks' (UNNs') outputs are combined in quadrature to provide a normalization factor for CP. Importantly, this method can be coupled with previously trained and optimized ML models. We apply this method to previously trained backbone neural networks (BNNs) that predict shielding factors in nuclear cross sections. Training, validation and testing data were calculated using Monte Carlo (MC) simulations of a light water reactor (LWR) pin cell [24]. The BNNs' input layer encodes the isotopic composition of uranium oxide (UOX) fuel for a LWR using scaled concentrations of 90 nuclides. The BNN includes two hidden layers, and an output layer consisting of 2,880 neurons representing shielding factors for 4 reactions, 8 energy groups and 90 nuclides. The uncertainty neural networks (UNNs) are trained using the activations calculated by the second hidden layer of the BNNs as input, along with the original ground truth training data. We find that our method produces estimates of both model and data uncertainty that are robust across BNN models whose mean relative errors (MREs) span approximately 1% to 0.1%. Our normalizer allows the predicted interval widths to be significantly narrower, simultaneously reduces the severity of overshoots dramatically, while maintaining the same level of coverage.

## 2. Background

This section is intended to provide a simple and high-level summary of CP methods. A deeper introduction to CP methods can be found in [17], and [25] provides a detailed and comprehensive mathematical treatment of CP and related methods.

### 2.1. Conformal Prediction.

The goal of CP is to compute an interval width $q$ such that the true values $y(x)$ of predictions $\hat{y}(x)$ fall inside an interval $C(x)$ with a probability of at least *1-*$\alpha$, Eq. (1):

$$\mathbb{P}\big(y(x) \in C(x)\big) \geq 1 - \alpha \qquad (1).$$

where $\alpha$ is a tunable parameter between 0 and 1 indicating the desired level of coverage, the interval $C(x)$ is defined by Eq. (2):

$$C(x) = [\hat{y}(x) - w, \hat{y}(x) + w] \quad (2).$$

and the $\hat{y}(x)$ predictions are calculated by a model using inputs *x*. The width *w* is calculated by ranking a set of nonconformity scores in ascending order. The nonconformity score measures how surprising the truth $y(x)$ is given a prediction $\hat{y}(x)$, and for point predictors the absolute residual $r(x)$ is a logical choice, Eq. (3):

$$s(x) = |r(x)| = |y(x) - \hat{y}(x)| \quad (3).$$

The probability of any new prediction $\hat{y}(x_{new})$ having a nonconformity score below a given threshold can then determined using the ranked set. The probability is given by the ratio of the number of data points in the set with nonconformity scores below the threshold, to the total number of data points. Typically in CP a desired level of coverage defined by $1\text{-}\alpha$ is sought, in which case the corresponding threshold is computed. For a ranked set of *n* data points and nonconformity score given by Eq. (3), the threshold is equivalent to the half-width $w_{1/2}$ and is given by Eq. (4):

$$w_{1/2} = q = Quantile\left(\{s(x_1), \dots, s(x_n)\}; \frac{(1-\alpha)(n+1)}{n}\right) \quad (4).$$

Equation (4) expresses that the width *w* is calculated by taking the empirical quantile of the nonconformity scores given by the target coverage level $1\text{-}\alpha$. In split conformal prediction (SCP) a calibration data set not used in training is held out and ranked to form the basis for calculating the interval width. Critically, to satisfy exchangeability all new data points must be drawn from the same distribution as the calibration data.

The CP method, however, can only guarantee that the calculated interval width *w* provides the specified level of coverage $1\text{-}\alpha$ on average across all new data points. This marginal coverage is expressed by Eq. (5):

$$\mathbb{P}\big(y(x_{new}) \in C(x_{new})\big) \geq 1 - \alpha \quad (5).$$

Conditional coverage, conversely, would hold for every specific data point:

$$\mathbb{P}(y(x_{new}) \in C(x_{new}) \mid x_{new} = x) \geq 1 - \alpha, \forall\, x \quad (6).$$

but has been shown to be impossible for a finite calibration set [20]. Consequently, the interval widths calculated using SCP can be excessively wide in some cases, and correspondingly narrow in others while meeting the nominal target coverage on average.

2.2. Adaptive Interval Widths.

The interval widths calculated using CP can be adjusted to mitigate imbalanced coverage by partitioning the calibration data into strata or categories. The categorization or stratification, however, must be done in terms of a quantity or categorical type whose ground truth is not required except during training and calibration. Mondrian conformal prediction (MCP) is a partitioning strategy where quantiles $q_k$ are calculated within subgroups $k$, Eq. (7):

$$q_k = Quantile\left(\{s(x_i)\ :\ x_i \in\ bin\ k\};\ \frac{(1-\alpha)(n_k+1)}{n_k}\right) \quad (7).$$

The coverage guarantee is now conditional on the subgroup $k$:

$$\mathbb{P}(y(x_{new}) \in C_k(x_{new}) \mid x_{new} \in\ bin\ k\ ) \geq 1-\alpha \quad (8).$$

Although by construction MCP calculates prediction intervals that guarantee coverage inside bins, these intervals remain flat within these subgroups. Computing intervals whose widths adapt continuously to the input can be done by scaling the nonconformity score, Eq. (3), using a normalization factor that is specific to a data sample $i$, Eq. (9).

$$s(x_i) = \frac{|y(x_i) - \hat{y}(x_i)|}{u_i} \quad (9).$$

Here the normalization factor $u_i$ is a quantity that estimates the difficulty of making a prediction for a sample $i$. Substituting nonconformity scores for a calibration dataset calculated using Eq. (9) into Eq. (7) then allows quantiles to be calculated. Prediction intervals for individual samples can then be found using Eq. (10), where the half-width $w_{1/2,k,i} = q_k u_i$:

$$C_k(x_i) = [\hat{y}(x_i) - q_k u_i, \hat{y}(x_i) + q_k u_i] \quad (10).$$

Although Eq. (10) is written for applicability to MCP, it is also applicable to CP without stratification, in which case the subscript $k$ is always equal to 1. In any case this approach, however, also requires the normalization factor $u_i$ to be accessible in unseen samples, where ground truth is unavailable.

## 3. Methods

3.1. Monte Carlo Simulations

The training data used in this work were generated using 10,000 OpenMC simulations of a 2-D light water reactor pin cell with reflective boundary conditions. Figure 1 shows the geometry and dimensions of the pin cell which consists of coolant, fuel, cladding, and a helium gap. The composition of the fuel included 90 nuclides and was varied in each simulation. The specific compositions were determined by first depleting fresh UOX fuel at enrichments of 1.6% to 19.9% a/o from 18 to 210 MWd/kgHM. This was done to establish the maximum and minimum expected concentrations of each of the 90 nuclides in diverse LWR fuel types. Latin hypercube sampling was then used to select a concentration (in atoms/b-cm) of each of the 90 nuclides, with an additional 10% added or subtracted to the maximum and minimum. A minimum concentration of zero was replaced with a trace amount of $10^{-21}$ a/o. All simulations were done using 10 inactive and 90 active criticality cycles with 10,000 neutrons per cycle. Full details of the data generation procedure and OpenMC simulations are provided in [24].

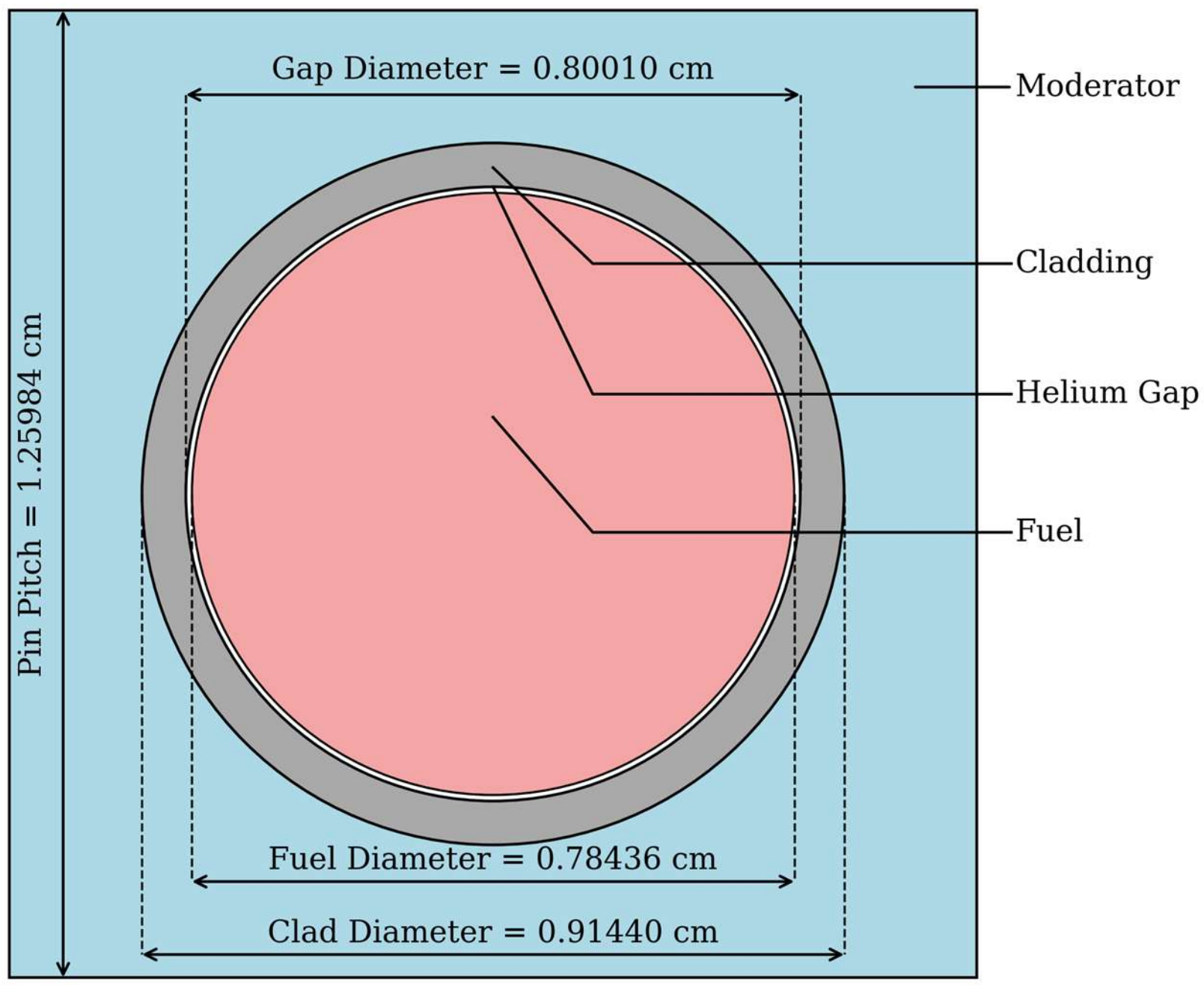

**Figure 1. Pin cell Geometry.** Periodic boundary conditions were applied to the enclosing perimeter around the coolant.

Each simulation tallied the integrated flux and the total, absorption, total scattering, and elastic scattering reaction rates for all 90 nuclides in the CASMO-8 group structure [26]. The MC standard deviations were also recorded for the flux and 4 reactions in each group. While these reactions were recorded for the fuel, cladding, helium gap, and water moderator, the neural networks in this work were trained using data derived from tallies generated only in the fuel. Microscopic cross sections were calculated by dividing the fuel tallies by the flux and nuclide concentrations. Tally uncertainties were calculated using standard error propagation and verified with the Python uncertainties library [27]. A histogram representing the relative uncertainties in all microscopic cross sections in the fuel is shown in Figure 2. The uncertainty was on average 0.43% of the ground truth value with a minimum value of 0.0837% and a maximum of 2.791%.

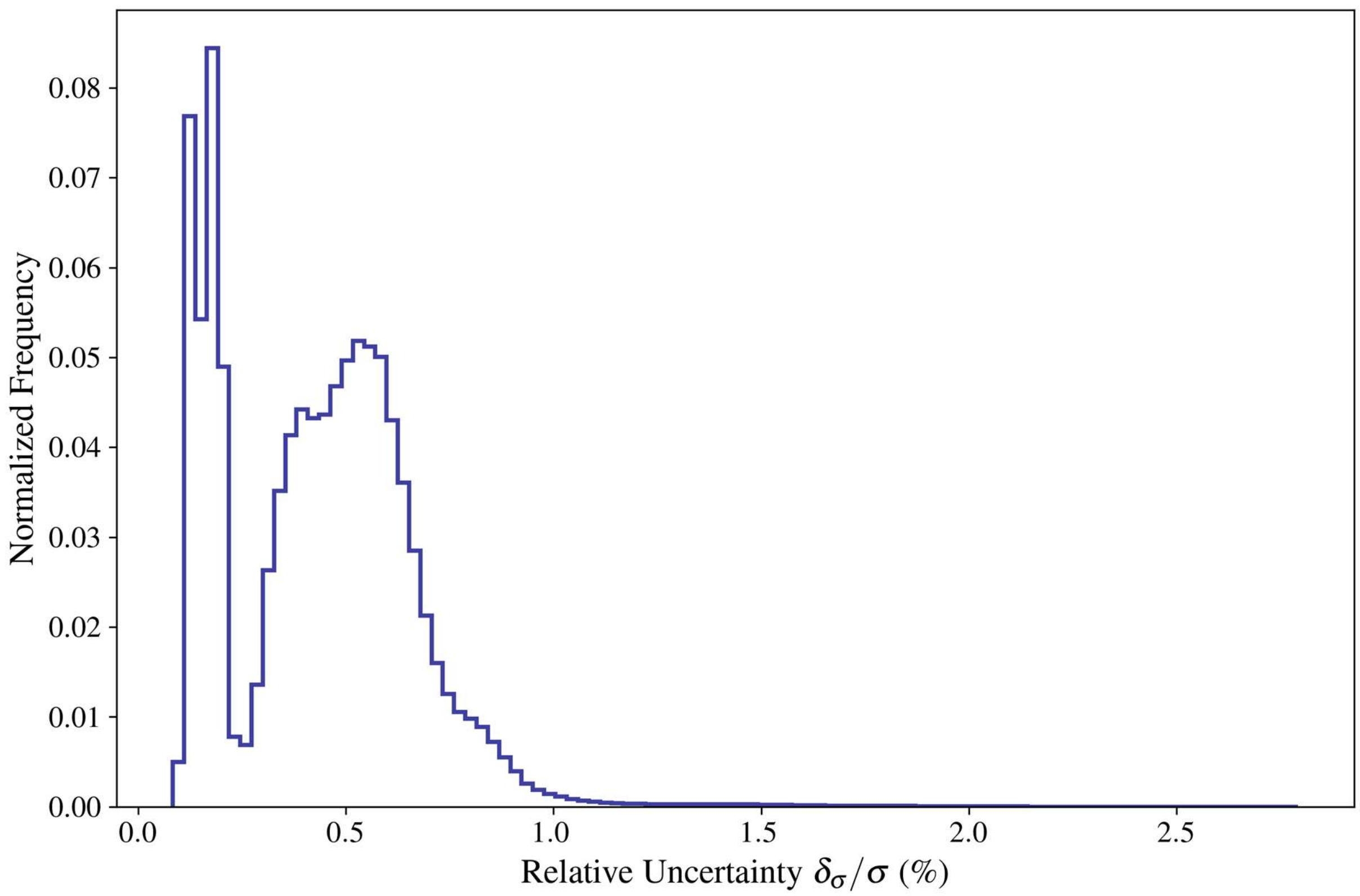

**Figure 2. Relative uncertainties of microscopic cross sections in the fuel.** The vertical axis shows the normalized frequency of all 28,800,000 data points (4 reactions, 8 groups, 90 nuclides, 10,000 samples). The horizontal axis shows the MC standard deviation divided by the microscopic cross section value. These data include total, absorption, total scattering and elastic scattering reactions in 90 nuclides in the fuel across 8 energy groups.

3.2. Training Data

Of the 10,000 training samples, 1,500 samples were set aside for testing and another 1,500 for validation and calibration. Each sample of training data consisted of the concentrations for all 90 nuclides included in the fuel as inputs and self-shielding ratios as outputs. The self-shielding ratios were calculated by dividing OpenMC-generated microscopic cross sections by the unshielded microscopic cross sections. The unshielded microscopic cross sections were calculated by taking the OpenMC ACE file data in continuous energy and collapsing the cross-sections into the CASMO-8 group structure using an arithmetic mean as shown in Eq. (11).

$$\sigma_g = \frac{1}{N_g} \sum_{i\,=\,E_g}^{E_{g+1}} \sigma\,(E_i) \qquad (11).$$

Where $N_g$ is the number of pointwise energy values within each CASMO-8 energy group $g$. Because the unshielded cross sections act only as a scaling factor for the output cross sections, the cross-section collapse does not require reaction rates to be conserved. Consequently, Eq. (11) was chosen due to being the simplest method. Figure 3 shows histograms of the cross-section ratios by reaction.

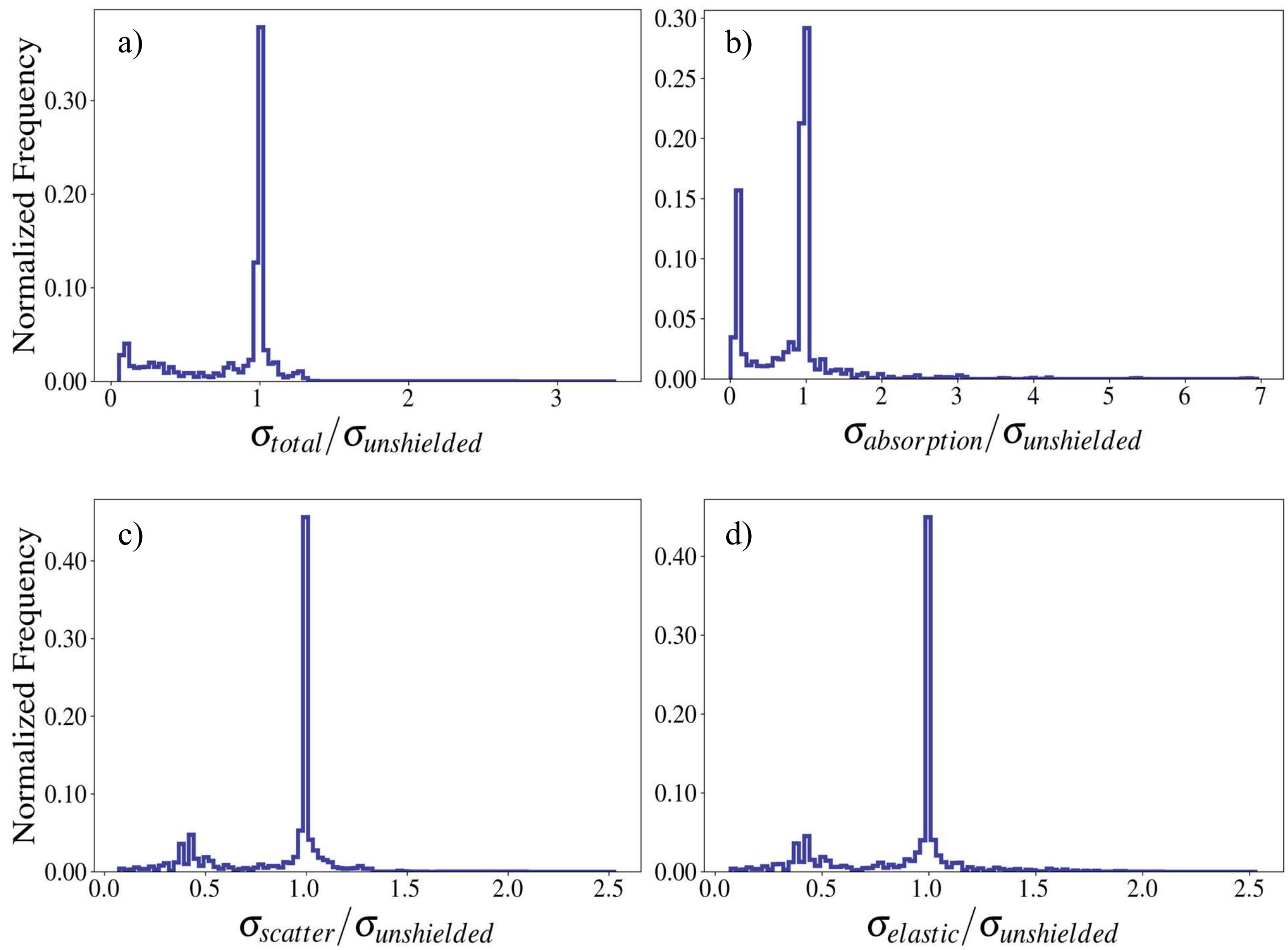

**Figure 3. Reaction cross-section ratios.** The histograms show the frequency of ratio values per reaction normalized by 7.2 million samples in each plot. Ratios are plotted by reaction, where a) shows values for the total, b) absorption, c) total scattering, and d) elastic scattering reactions.

In each reaction shown in Figure 3, the majority of data points have very little to no self-shielding, as the bin closest to one has the highest frequency. Values below one are shielded effective cross sections while values above one represent an increase in effective cross section.

The uncertainties in the cross-section ratios were also calculated using standard error propagation and validated using the Python uncertainties library. Although nuclear data libraries include covariance data, our analysis did not include uncertainties in the ENDF data. Relative uncertainties in the cross-section ratios are shown in Fig. 4 for each reaction.

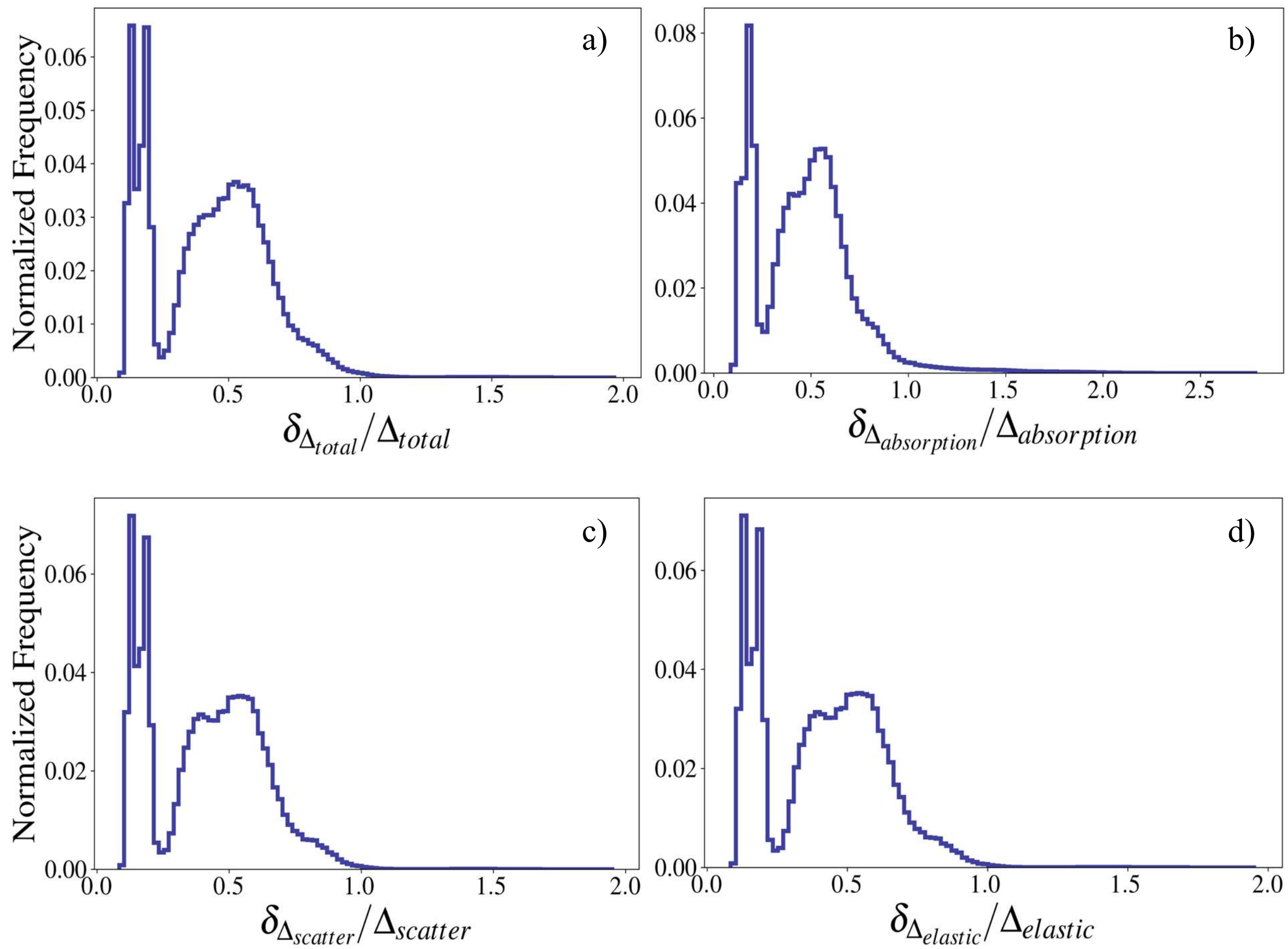

**Figure 4. Relative uncertainties in the self-shielding ratios.** These histograms show the frequency of relative uncertainty values ($\delta_\Delta$) for the self-shielding ratios ($\Delta$) per reaction normalized by the 7.2 million samples in each plot. Uncertainties are plotted by reaction, where a) shows values for the total, b) absorption, c) total scattering, and d) elastic scattering reactions

The relative uncertainty distributions shown above look similar to those in Figure 2 because uncertainties on the self-shielding ratios are solely from MC noise. The relative uncertainty does not exceed 2.79% of the ratio values for any reaction, which is consistent with the uncertainty seen on the microscopic cross sections generated by OpenMC.

3.3. Neural Networks

3.3.1. Backbone neural networks

Backbone Neural Networks (BNNs) were pre-trained models that estimated shielding ratios, onto which uncertainty predicting neural networks would be attached. The neural network trained and described in [24] was adapted in this work as the pre-trained target for testing and application of our uncertainty methodology. The network accepted the 90 scaled nuclide concentrations as input and predicted cross-section ratios. The network included two hidden layers, and an output layer consisting of 2,880 neurons, each corresponding to a nuclide, energy group index and reaction type. The inputs were scaled using the SciPy standard scaler [28], while the outputs were left unscaled as the ground truth values were ratios ranging from 0.0001 to 14. The mean relative error (MRE) for the test data of the original network in [24] was approximately 0.15%. Consequently, to test the applicability of our uncertainty method to less accurate models, the network was re-trained five times with identical hyperparameters and topology. Training runs were configured to terminate once the validation MRE reached targets of 0.1%, 0.25%, 0.5%, 0.75% and 1%, or when 5000 epochs were completed. To ensure consistency the original network was also re-trained, and the test MRE values reached in each model's training run were 0.125%, 0.248%, 0.482%, 0.725% and 0.987%. The most accurate model reached 5000 epochs during training, which terminated before the 0.1% validation MRE was reached. In the training of the remaining models, from most to least accurate the training runs completed 421, 98, 61 and 44 epochs before terminating. For clarity and ease of comparison, the five BNNs' key training characteristics given above are also listed in Table I below.

**Table I. Backbone Neural Network Training Characteristics**

| BNN | Target Validation MRE | Test MRE Reached | Training Epochs Completed |
|---|---|---|---|
| BNN-0.1 | 0.1% | 0.125% | 5000 |
| BNN-0.25 | 0.25% | 0.248% | 421 |
| BNN-0.5 | 0.5% | 0.482% | 98 |
| BNN-0.75 | 0.75% | 0.725% | 61 |
| BNN-1.0 | 1.0% | 0.987% | 44 |

In all five BNNs, the loss function used during training is given by Eq. (12) [29]. The loss term is based on a negative log-likelihood function, intended to reduce the penalty for less accurate predictions when the ground truth uncertainty is large.

$$\mathcal{L}_{bb} = \sum_i \frac{1}{2}\left[\frac{[y(x_i) - \hat{y}(x_i)]^2}{max(\sigma_{omc}(x_i), \epsilon)} + \log\left(max(\sigma_{omc}(x_i), \epsilon)\right)\right] \qquad (12).$$

Here, $\sigma_{omc}(x_i)$ is the ground-truth standard deviation on the cross-section ratios calculated as described in Section 3.2. The symbol $\epsilon$ is the PyTorch default value of $1 \times 10^{-6}$ to clamp the result of the calculation. The topology and hyperparameters of the BNNs are given in Table I.

3.3.2. Uncertainty neural networks

The residual of predictions made with the backbone neural networks, Eq. (3), has an expectation value that can be decomposed into model and statistical variances. These are intrinsic to the BNN model, and related to the stochastic nature of the Monte Carlo calculation respectively:

$$\mathbb{E}[r^2 \mid x] = \delta(x)^2 + [\sigma_{omc}(x)]^2 \tag{13}$$

Here $[\sigma_{omc}(x)]^2$ is the variance in cross-section ratios from tallies calculated by OpenMC, and $\delta(x)^2$ is the variance due to the backbone model's approximation of the ground truth. Because neither of these quantities are available for new samples, uncertainty neural networks (UNNs) were trained to predict each.

For each of the five previously trained BNNs, a $\delta$-predicting UNN and $\sigma$-predicting UNN were trained, giving a total of ten UNNs. The $\delta$-UNNs were trained to predict their corresponding BNN's model error, while the $\sigma$-UNNs were trained to predict the MC statistical error, for individual input samples in both cases. UNNs' predictions consisted of 2880 uncertainties that corresponded to the BNNs' cross section ratios. During the training of the UNNs, the weights and biases of the BNNs remained fixed.

The inputs to the UNNs were the activations calculated by the BNN's penultimate layer, which varied according to the input isotopic compositions. Figure 5 illustrates how the UNNs were connected to the BNNs. The hyperparameters used in training both UNNs are given in Table II. Equations (14) and (15) represent the loss functions used to train the $\delta$-UNNs and $\sigma$-UNNs respectively. Equation (14) represents the loss function used to train the $\delta$-UNNs:

$$\mathcal{L}_\delta = \sum_i \left[ \frac{[y(x_i) - \hat{y}(x_i)]^2}{2\left(\hat{\delta}(x_i)^2 + [\sigma_{omc}(x_i)]^2\right)} + \frac{1}{2} \log\left(\hat{\delta}(x_i)^2 + [\sigma_{omc}(x_i)]^2\right) \right] \tag{14}$$

The $\delta$-UNNs emit a log-variance $\ell_\delta$ which forces the predicted model variance $\hat{\delta}(x_i)^2 = e^{\ell_\delta}$ to be positive. Equation (15) represents the loss function used to train the $\sigma$-UNNs:

$$\mathcal{L}_\sigma = \sum_i \left[ 2\log\left(\hat{\sigma}(x_i)\right) - \log\left([\sigma_{omc}(x_i)]^2 + \epsilon\right) \right]^2 \tag{15}$$

The $\sigma$-UNNs also emit a log-variance $\ell_\sigma$ which forces the predicted MC standard deviation $\hat{\sigma}(x_i) = e^{\frac{1}{2}\ell_\sigma}$ to be positive. The $\epsilon$ symbol is a small epsilon added to avoid infinities if $\sigma_{omc}(x_i)$ is zero. The quantity $\sigma_{omc}(x_i)$, present in both equations (14) and (15), is the same ground-truth standard deviation in the cross-section ratios as in Eq. (13). Consequently, data used to train these networks must include uncertainties. The hyperparameters and topology of the UNNs is given in Table II.

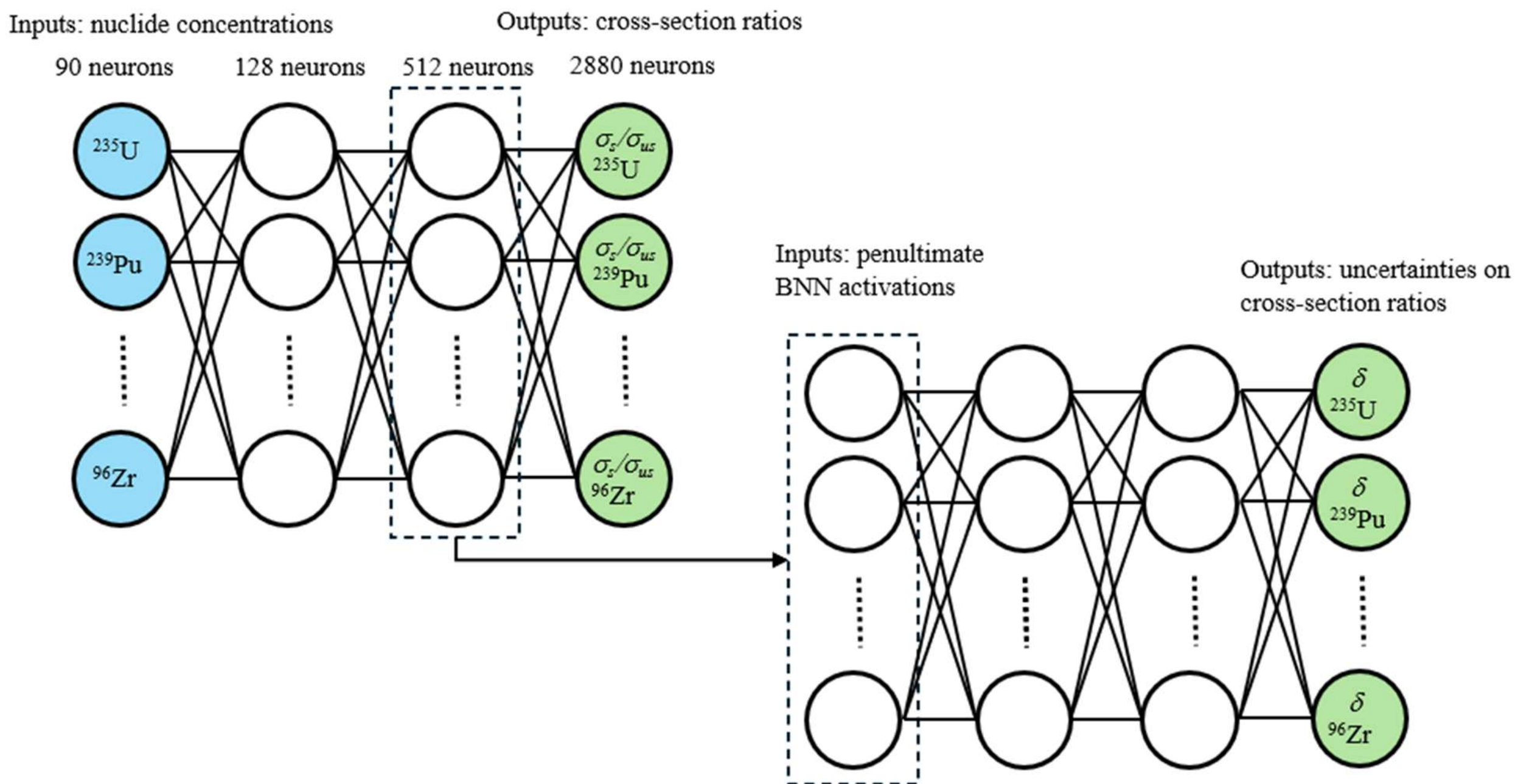

**Figure 5. Topology of the combined BNN + $\boldsymbol{\delta}$-UNN networks**. The topology of the BNN + $\sigma$-UNNs is identical. Although the $\delta$-UNNs and $\sigma$-UNNs were trained independently of each other, both networks could be connected to the BNNs in parallel.

**Table II. Network Hyperparameters and Topology**

| | **BNNs** | | | | **UNNs** | | | |
|---|---|---|---|---|---|---|---|---|
| **Optimizer** | Adam | | | | Adam | | | |
| **Batch Size** | 300 | | | | 300 | | | |
| **Epochs** | 5000 | | | | 5000 (maximum) | | | |
| **Learning Rate** | 0.001 | | | | 0.001 | | | |
| **Learning Rate Scheduler** | Reduce LR on Plateau Patience = 250, Factor = 0.5 | | | | Constant LR | | | |
| **Weight Decay** | 0 | | | | $10^{-5}$ | | | |
| **Dropout** | None | | | | None | | | |
| **Layer Index** | **1** | **2** | **3** | **4** | **1** | **2** | **3** | **4** |
| **Topology (Neurons in Layer)** | 90 | 128 | 512 | 2880 | 512 | 64 | 64 | 2880 |
| **Activation Function** | n/a | ReLU | ReLU | n/a | n/a | SiLU[a] | SiLU[a] | n/a |

a. SiLU activation function refers to Sigmoid Linear Unit.

## 4. Results and discussion.

### 4.1 Monte Carlo and Model Uncertainties

The reliability curves given in Fig. 6 show the MC and model uncertainties predicted by the $\sigma$-UNNs and $\delta$-UNNs for BNNs in order of increasing BNN accuracy. The empirical model error cannot be extracted from individual predictions but can be estimated by binning according to similarity in their predicted model error $\hat{\delta}$. For a bin $B$ with $N_B$ samples it can be shown that the empirical model error in bin $B$ is given by:

$$\delta_B^2 = max(0, \langle r^2 \rangle - \langle \sigma_{omc}^2 \rangle) \quad (16).$$

Justification for Eq. (16) is provided in Appendix A. Consequently, the reliability curves for the $\delta$-UNNs exclude samples where the empirical model error is smaller than the empirical MC error. The reliability curves show that in each equal-frequency bin both the predicted model and MC uncertainties on average align extremely well with the empirical uncertainties. At all levels of BNN accuracy the $\sigma$-UNNs' MC error predictions are similar, as expected given that the uncertainty in the ground truth values doesn't change as the backbone MRE improves. The drop in the $\delta$-UNNs' reliability curves at low model error levels is due to the empirical MC uncertainty becoming comparable to the empirical model uncertainty. This makes it impossible to separate the empirical model and MC error by subtracting the empirical MC error from the residual.

The reliability diagrams shown in Fig. 6 only provide an indication of the UNNs' reliability in an average sense over bins in the predicted quantities. In a well-trained uncertainty model, the residuals from the BNN's predictions should be close in size to the combined model and MC uncertainties. We quantify this using the standardized residual score, defined by Eq. (17) for a prediction $i$:

$$z_i = \frac{y(x_i) - \hat{y}(x_i)}{\sqrt{\left(\hat{\delta}(x_i)^2 \ + \ [\sigma_{omc}(x_i)]^2\right)}} \quad (17).$$

The numerator of Eq. (17) is the BNN's signed residual, while the denominator is the predicted one-sigma expected spread of the residual. We test the accuracy of the $\delta$-UNN in isolation by using the predicted model uncertainty $\hat{\delta}(x_i)$ and the empirical MC uncertainty $\sigma_{omc}(x_i)$ in the denominator of Eq. (17). Figure 7 shows the distribution of the standardized residual score calculated using individual predictions made by the $\delta$-UNNs. Because the residuals from BNN-0.1, BNN-0.25 and BNN-0.5 are comparable to the empirical MC uncertainty, standardized residuals in the model and MC uncertainty dominated region are plotted separately.

The summary statistics of the histograms shown in Fig. 7 are also given in Table III. For a well-calibrated predictor of the model uncertainty the resulting standardized residual distribution should follow:

$$z \sim \mathcal{N}(0,1) \quad (18).$$

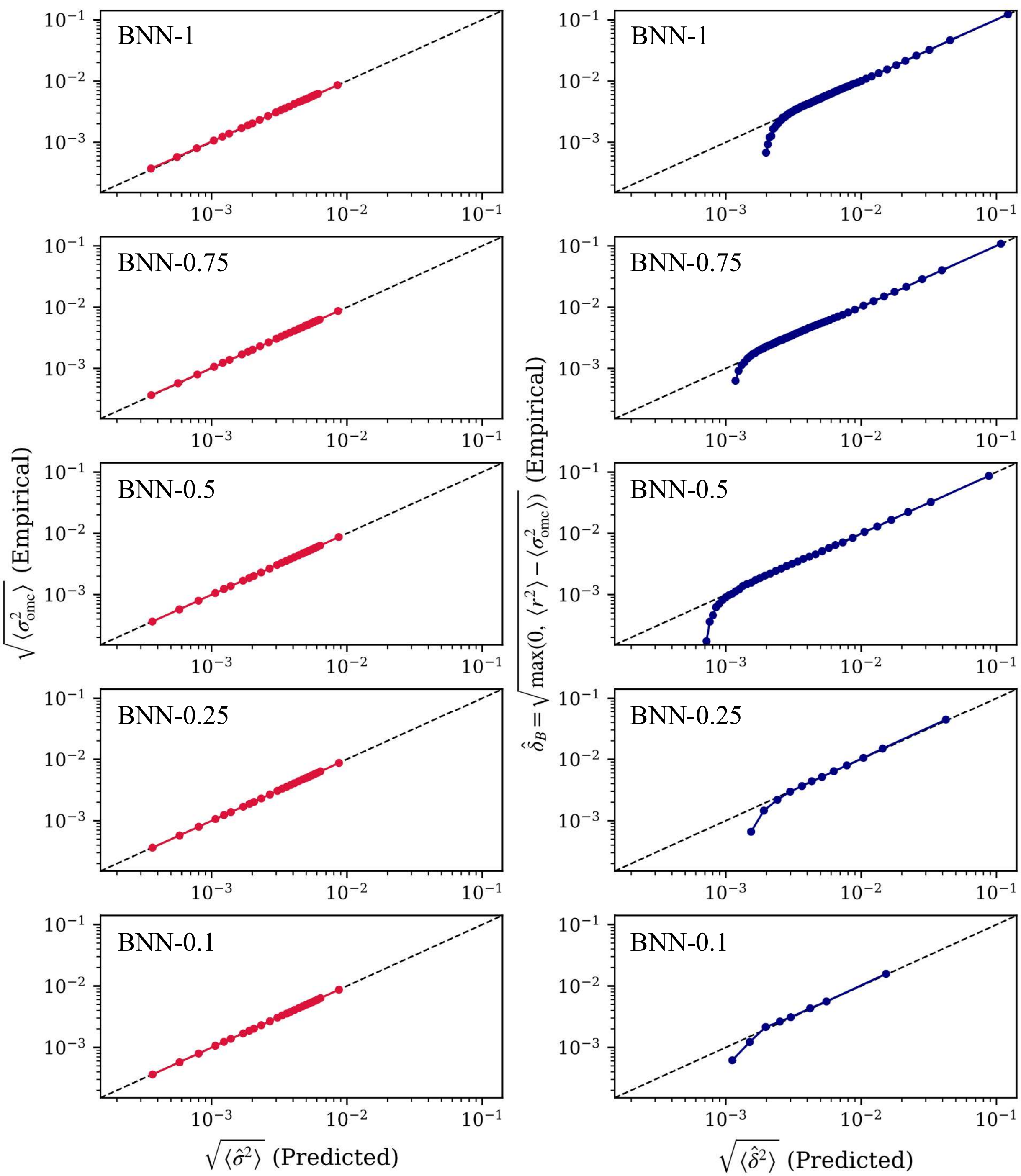

**Figure 6. Reliability diagrams comparing empirical and predicted MC ($\sigma$-UNNs, left column) and model ($\delta$-UNNs, right column) uncertainties, test data.** From top to bottom each pair of plots corresponds to backbone models with target MRE values from 1% to 0.1%. The MC errors are shown in 25 equal-frequency bins, binned according to the predicted MC error. The model errors are shown in up to 100 equal-frequency bins, with bins excluded when the empirical MC error is smaller than the empirical model error.

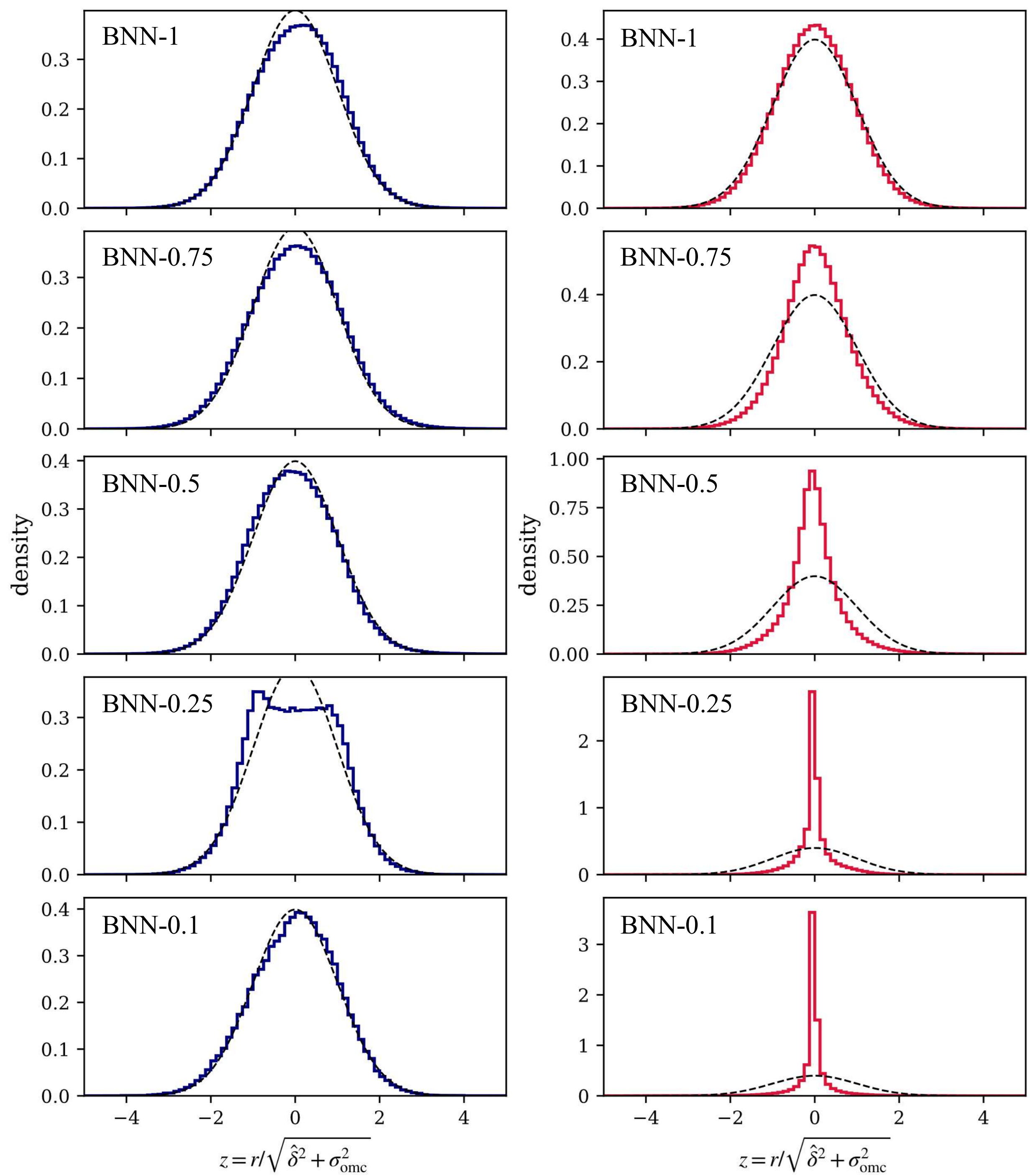

**Figure 7. Standardized residuals calculated using $\delta$-UNN predicted model uncertainties, test data.** The left column includes only samples whose predicted model uncertainty $\hat{\delta}$ is greater than the OpenMC MC uncertainty $\sigma_{omc}$. The column on the right includes only the remaining samples. The dashed curves follow a normal distribution with zero mean and standard deviation of unity. From top to bottom each pair of plots corresponds to backbone models with target MRE values from 1% to 0.1%.

**Table III. Figure 7 Summary Statistics.**

| Target MRE | Model Uncertainty Dominated | | MC Uncertainty Dominated | |
|---|---|---|---|---|
| | Fraction of samples | Histogram Standard Deviation | Fraction of samples | Histogram Standard Deviation |
| 1.0% | 64.3% | 1.03 | 35.7% | 0.93 |
| 0.75% | 49.6% | 1.08 | 50.4% | 0.82 |
| 0.5% | 26.9% | 1.03 | 73.1% | 0.69 |
| 0.25% | 11.0% | 1.02 | 89.0% | 0.46 |
| 0.1% | 4.6% | 1.02 | 95.4% | 0.43 |

These results show that all $\delta$-UNNs are well-calibrated in the region where the model uncertainty dominates. In the MC uncertainty dominated region, only $\delta$-UNN-1.0 is well-calibrated. This result is due to the way in which the built-in OpenMC tally arithmetic functions propagate uncertainty [30]. The microscopic cross-sections are calculated by dividing reaction rate and flux tallies:

$$\sigma = \frac{1}{N}\frac{RR}{\phi} \tag{19}$$

where $RR$ is the reaction rate [1/cm$^3$-s], $N$ is the atomic concentration [1/cm$^3$] and $\phi$ is the neutron flux [1/cm$^2$-s]. Because both the reaction rate and the flux are scored through the same neutron tracks, the numerator and denominator of Eq. (19) are correlated. OpenMC tally arithmetic, however, propagates uncertainty by assuming the two tallies are independent. While this produces conservative estimates of uncertainty, the effect is that the uncertainties in the ground truth data used to train the NNs are inflated.

Figure 7 also clearly shows that the normalized residual calculated using $\delta$-UNN-0.25 in the model error dominated region is not normally distributed. While all the $\delta$-UNNs' normalized residuals are not precisely normally distributed, this is mainly because the residuals cluster near values similar in magnitude to the denominator of Eq. (17). This effect is most pronounced in the case of $\delta$-UNN-0.25 where the distributions of the residuals, predicted model error and MC error have a greater overlap.

Figure 8 shows the relative residuals of the MC uncertainties predicted by the $\sigma$-UNNs. The results are also separated according to whether the predictions fall in the region dominated by predicted model and MC error. The $\sigma$-UNNs recover the reported $\sigma_{omc}$ to roughly 10%, with a small BNN-dependent median offset of a few percent. Table IV gives the summary statistics of the histograms shown in Fig. 8. The fraction of data samples in each split is the same as given in Table III.

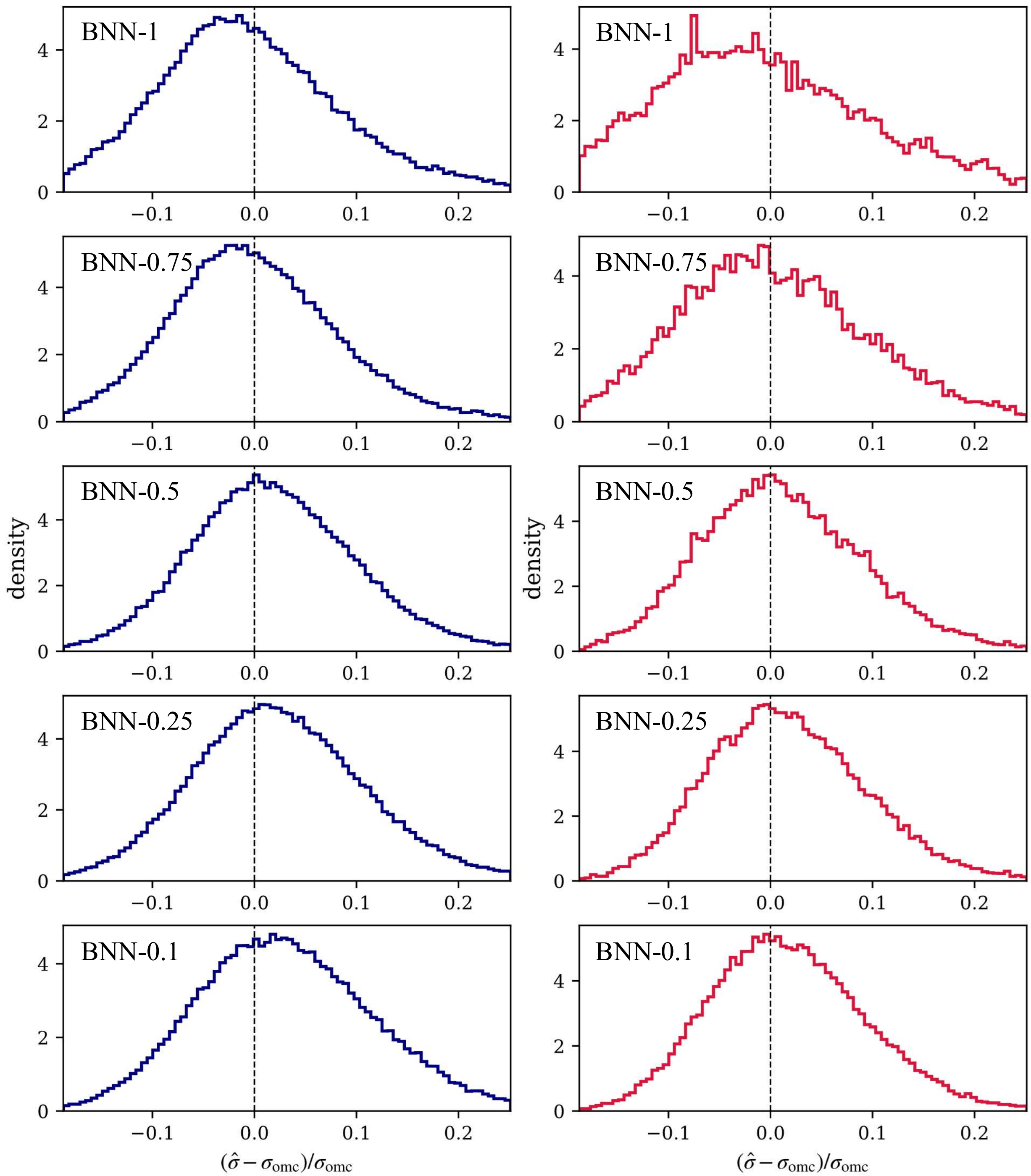

**Figure 8. Relative residuals calculated using $\sigma$-UNN predicted** MC **uncertainties, test data.** The left column includes only samples whose predicted model uncertainty $\hat{\delta}$ is greater than the OpenMC uncertainty $\sigma_{omc}$. The column on the right includes only the remaining samples. From top to bottom each pair of plots corresponds to backbone models with target MRE values from 1% to 0.1%.

**Table IV. Figure 8 Summary Statistics.**

| | **Model Uncertainty Dominated** | | **MC Uncertainty Dominated** | |
|---|---|---|---|---|
| Target MRE | Median | Histogram Standard Deviation | Median | Histogram Standard Deviation |
| 1.0% | -0.013 | 0.103 | -0.018 | 0.119 |
| 0.75% | -0.006 | 0.088 | -0.004 | 0.097 |
| 0.5% | 0.016 | 0.086 | 0.007 | 0.081 |
| 0.25% | 0.021 | 0.087 | 0.012 | 0.078 |
| 0.1% | 0.027 | 0.089 | 0.013 | 0.078 |

4.2 Conformal Prediction

The outputs of the $\delta$-UNN and $\sigma$-UNN can be combined according to Eq. (20) to provide a normalization factor $u_i$ for Eqs. (9,10):

$$u_i = \sqrt{\left(\hat{\delta}(x_i)^2 \ + \ [\sigma_{omc}(x_i)]^2\right)} \qquad (20).$$

In this section we use un-normalized, i.e. $u_i = 1$, and unstratified, i.e. Eqs. (1-5) CP as a baseline against which our normalization method can be compared. Importantly, although the unstratified CP method does not partition the data, results are binned for visualization only.

Figures 9 and 10 show the coverage as a function of the binned self-shielding strength for a target coverage of 99% and 95% respectively. Each data point represents the mean coverage level in each equal-frequency bin. In these plots data points are often clustered together horizontally, which is an effect of equal-frequency binning. The results show that both un-normalized and normalized CP consistently over-cover the prediction when the shielding is weak, while under-covering in the strong shielding region. An ideal level of coverage would track the target coverage exactly, but in unstratified CP the coverage guarantee holds only on average across all predictions. Consequently, over-coverage in a subset of the predictions must be balanced by under-coverage elsewhere. The results also show, however, that using the normalization of Eq. (20) partially restores coverage in the high shielding region. The recovery is more pronounced in the results corresponding to BNNs with MRE values of 0.25% and greater.

Figures 11 and 12 show the interval widths as a function of the binned self-shielding strength for a target coverage of 99% and 95% respectively. In the un-normalized CP results, the widths are necessarily constant because a single width is chosen to provide the target coverage over the entire data. The results also show that using normalization dramatically reduces the predicted interval widths for the 99% coverage target. For the 95% target, results are more mixed, showing clearer improvement only when the BNNs MRE values are 0.5% or greater.

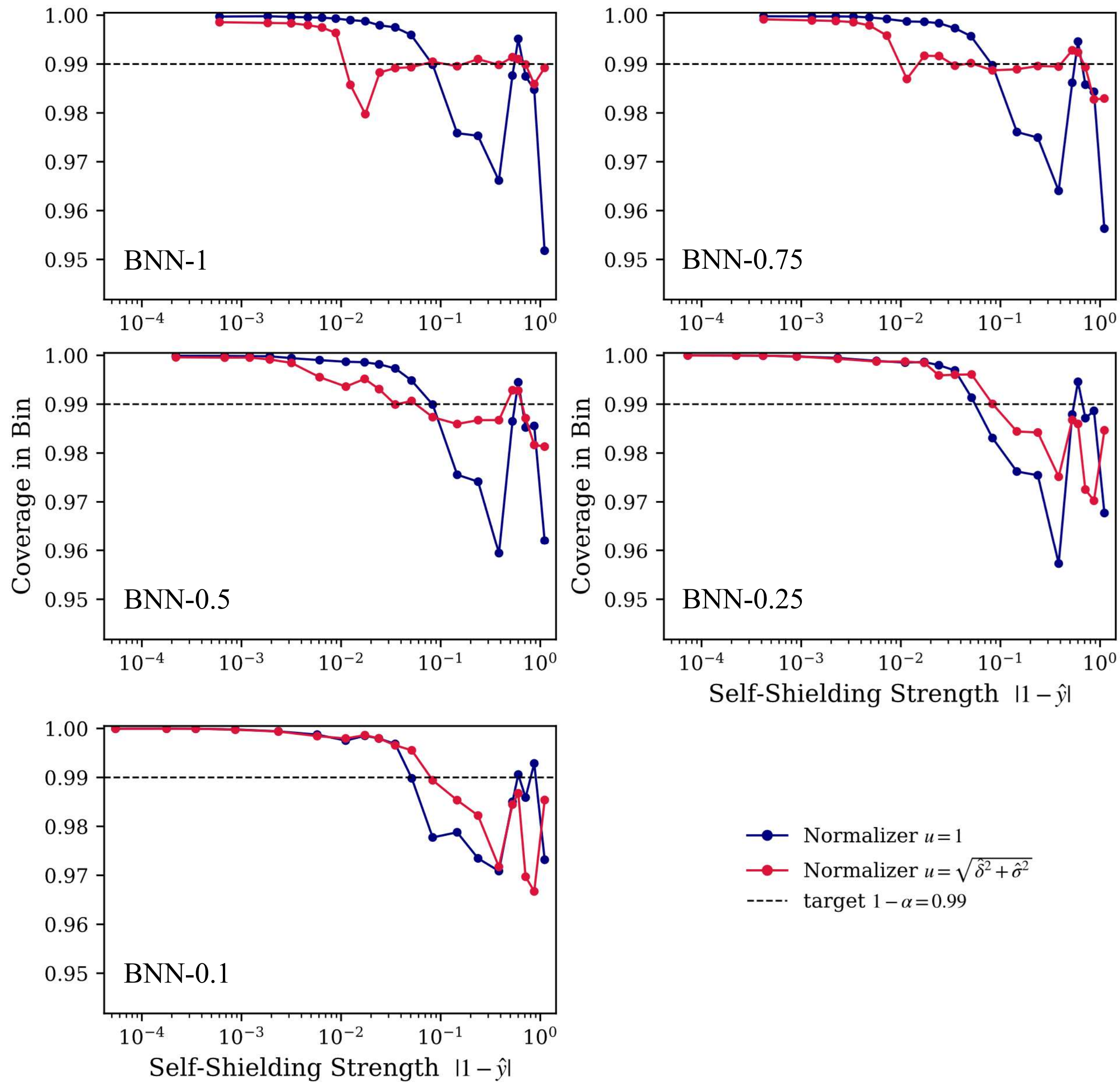

**Figure 9. Coverage level for 99% target, unstratified conformal prediction.** Individual plots correspond to BNNs with increasing levels of accuracy. The coverage calculated using un-normalized CP, Eq. (2) and normalized CP, Eqs. (10,20) with $k=1$, are shown by the blue and red curves respectively. The dashed line represents the target coverage level. The self-shielding strength is greatest at the right-hand side of each plot, corresponding to small predicted cross-section ratios. Binning here is done for visualization only.

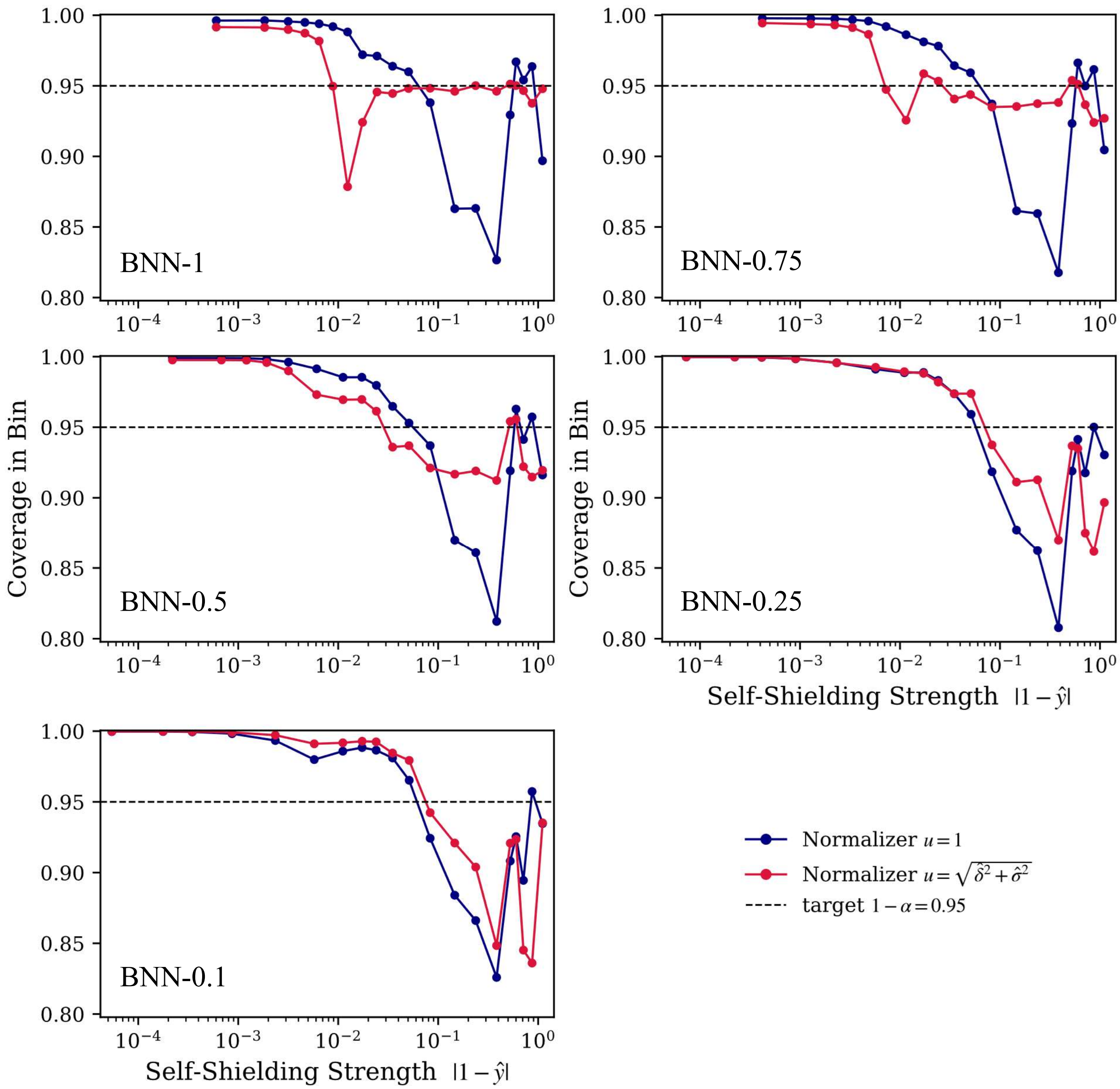

**Figure 10. Coverage level for 95% target, unstratified conformal prediction.** Individual plots correspond to BNNs with increasing levels of accuracy. The coverage calculated using un-normalized CP, Eq. (2) and normalized CP, Eqs. (10,20) with *k=1,* are shown by the blue and red curves respectively. The dashed line represents the target coverage level. Binning here is done for visualization only.

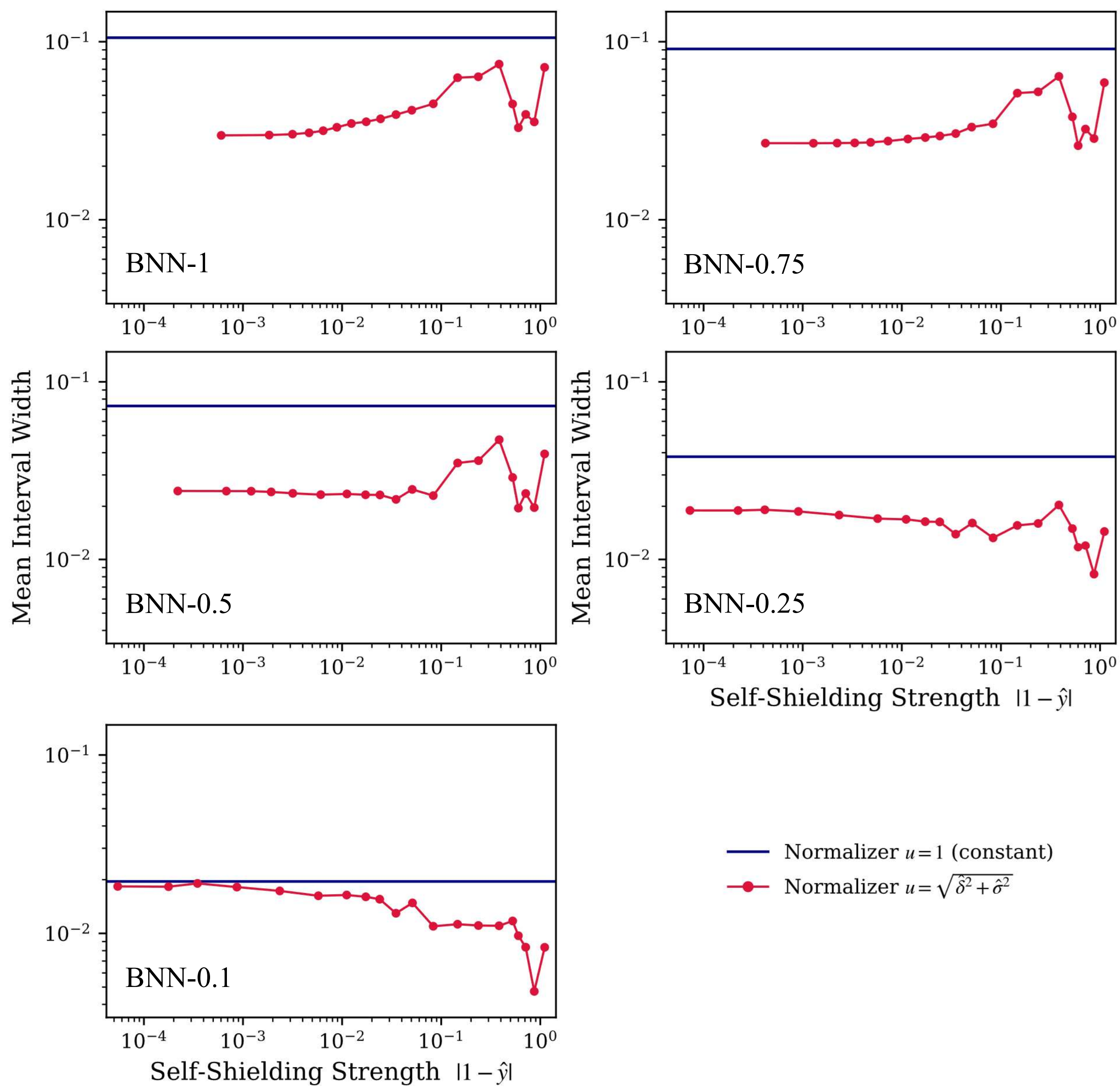

**Figure 11. Predicted interval widths for 99% target, unstratified conformal prediction.** Individual plots correspond to BNNs with increasing levels of accuracy. The widths $w$ calculated using un-normalized CP, Eq. (4), and normalized CP, Eqs. (10,20) with $k=1$, are shown by the blue and red curves respectively. Because the normalizer adjusts widths for individual samples, the widths shown in the normalized red curve represent the mean width in the self-shielding strength bin. Binning here is done for visualization only.

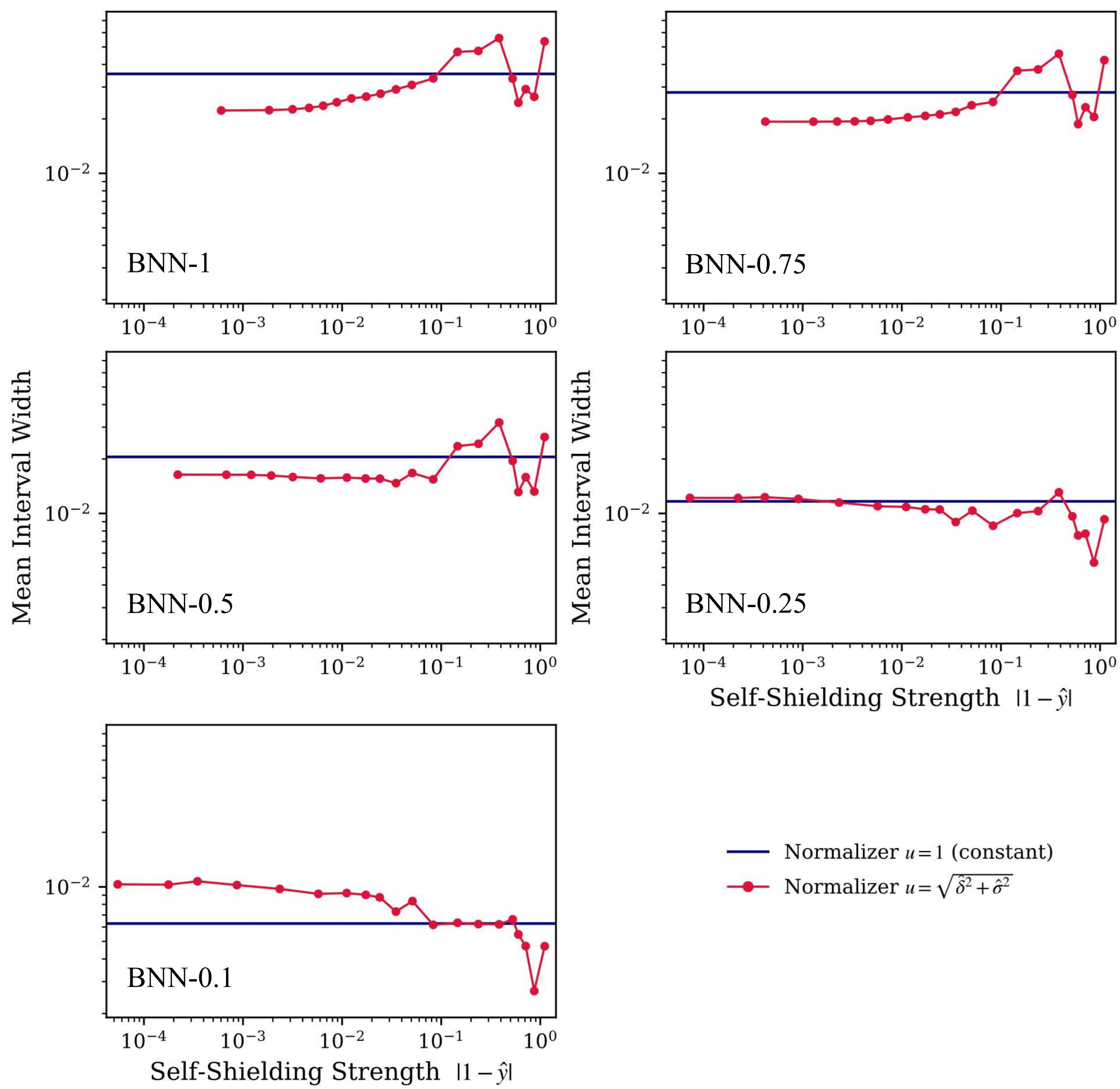

**Figure 12. Predicted interval widths for 95% target, unstratified conformal prediction.** Individual plots correspond to BNNs with increasing levels of accuracy. The widths $w$ calculated using un-normalized CP, Eq. (4), and normalized CP, Eqs. (10,20) with $k=1$, are shown by the blue and red curves respectively. Because the normalizer adjusts widths for individual samples, the widths shown in the normalized red curve represent the mean width in the self-shielding strength bin. Binning here is done for visualization only.

### 4.3 Stratified Conformal Prediction: Coverage and Interval Widths

In this section we show coverage levels and predicted interval widths using un-normalized, i.e. $u_i = 1$, and normalized, i.e. Eq. (20), CP where both are combined with stratification according to Eqs. (7,8). We choose the two most salient stratification axes that are accessible when no ground truth is available to show our results. The first axis we choose is the predicted self-shielding strength, which is provided by the BNNs, and is the abscissa in Figs. 9-12. Secondly and independently, we stratify the data according to the specific output channel of the BNNs. These outputs are the cross-section ratios of 4 reactions, 90 isotopes and 8 energy groups giving a total of 2,880 strata. Because the calibration data set consists of 1,500 samples, each stratum contains sufficient samples to ensure that the partitioning, Eq. (7), yields meaningful quantiles in a statistical sense.

#### 4.3.1 Stratification by predicted self-shielding strength

Figures 13 and 14 show the coverage as a function of the binned self-shielding strength for a target coverage of 99% and 95% respectively. Because the predictions are stratified and plotted along the same axis, the coverage is dramatically more balanced than unstratified CP, Figs. 9,10. Although the results show that the target coverage levels are met, this should not be interpreted as a success of either method, as marginal coverage is guaranteed by construction. Because the CP prediction guarantee is in an average sense, individual data points sitting at coverage levels slightly higher or lower than the target are expected. For the same reason, little effective difference in coverage between the normalized and un-normalized CP is also expected.

Figures 13 and 14 also suggest that the coverage of both CP methods trends slightly higher than the target, particularly for the higher MRE BNNs. This is also expected as the coverage condition, Eq. (8), seeks interval widths that equal or exceed the target coverage level. Figures 15 and 16 show the mean interval width in each self-shielding strength bin for target coverage levels of 99% and 95% respectively. These results show that with few exceptions, the normalization approach dramatically reduces the predicted interval widths when self-shielding is significant. This is particularly pronounced at a 99% coverage target, Fig. 13, where the normalizer increases predicted widths only slightly for BNN MRE levels below 0.5% when self-shielding is weak. In these cases the widths are already much narrower than at high self-shielding levels, despite the magnitude of the BNN prediction $\hat{y}$ being larger. At 95% coverage the trend remains although the effect is not as pronounced. Results from Figs. 15 and 16 are summarized in Table V.

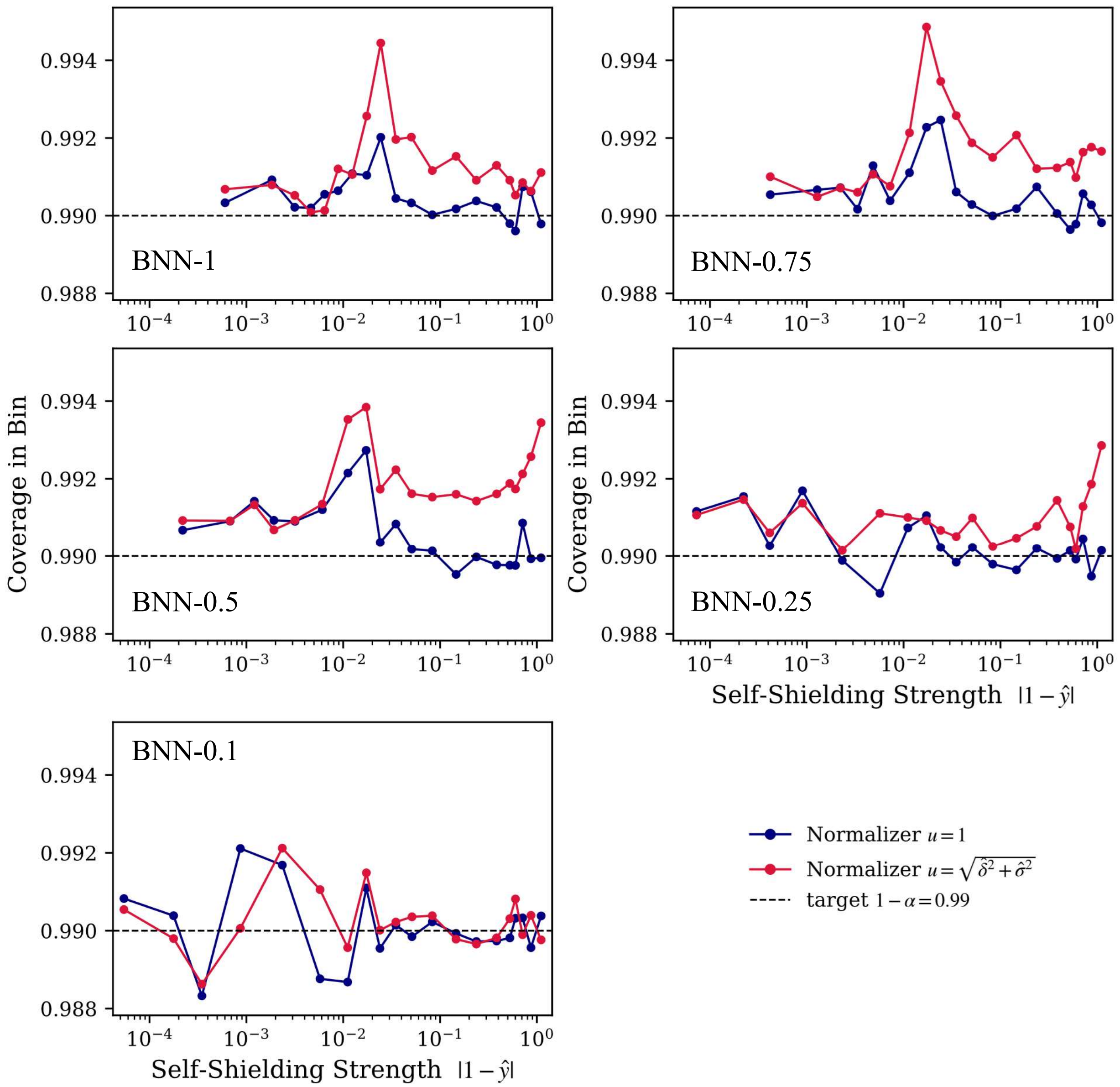

**Figure 13. Coverage level for 99% target, CP stratified by self-shielding strength.** Individual plots correspond to BNNs with increasing levels of accuracy. The coverage calculated using un-normalized CP, Eq. (7) and normalized CP, Eq. (10), are shown by the blue and red curves respectively. The dashed line represents the target coverage level. The self-shielding strength is greatest at the right-hand side of each plot, corresponding to small predicted cross-section ratios.

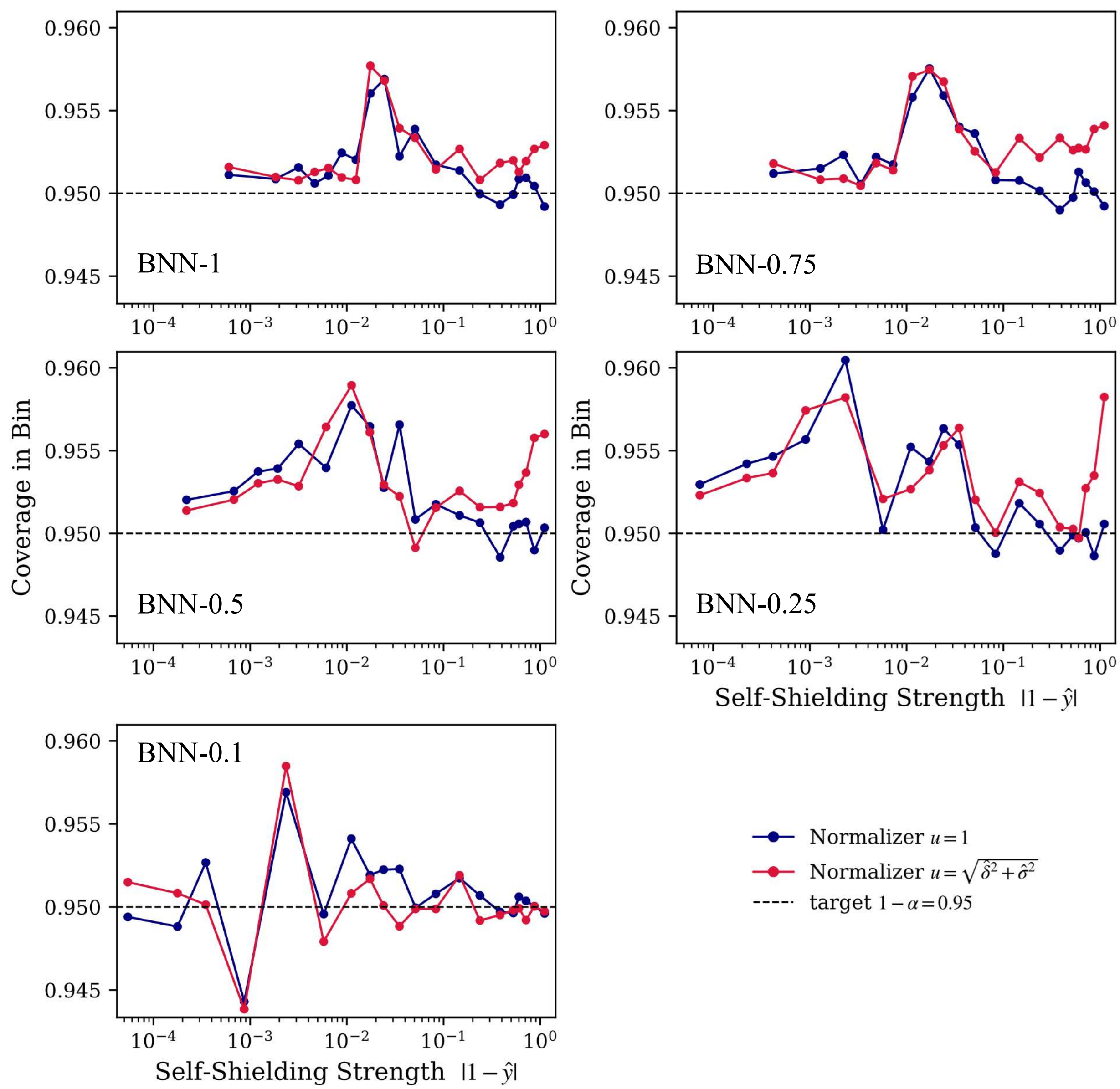

**Figure 14. Coverage level for 95% target, CP stratified by self-shielding strength.** Individual plots correspond to BNNs with increasing levels of accuracy. The coverage calculated using un-normalized CP, Eq. (7) and normalized CP, Eq. (10), are shown by the blue and red curves respectively. The dashed line represents the target coverage level. The self-shielding strength is greatest at the right-hand side of each plot, corresponding to small predicted cross-section ratios.

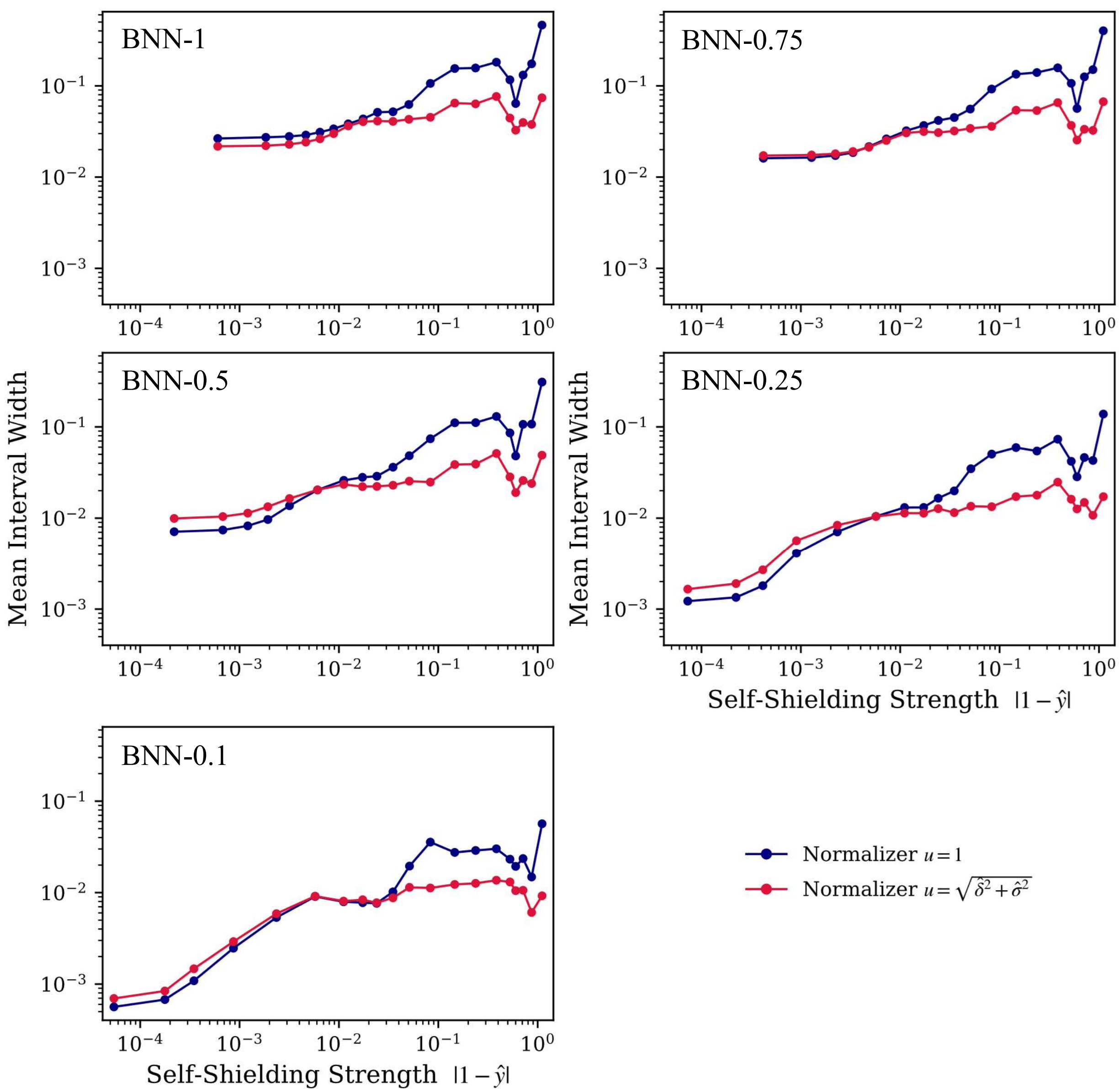

**Figure 15. Predicted interval widths for 99% target, CP stratified by self-shielding strength.** Individual plots correspond to BNNs with increasing levels of accuracy. The widths $w$ calculated using un-normalized CP, Eq. (7) and normalized CP, Eq. (10), are shown by the blue and red curves respectively. The self-shielding strength is greatest at the right-hand side of each plot, corresponding to small predicted cross-section ratios. Because the normalizer adjusts widths for individual samples, the widths shown in the normalized red curve represent the mean width in the self-shielding strength bin. The widths corresponding to the un-normalized blue curve are constant in each bin.

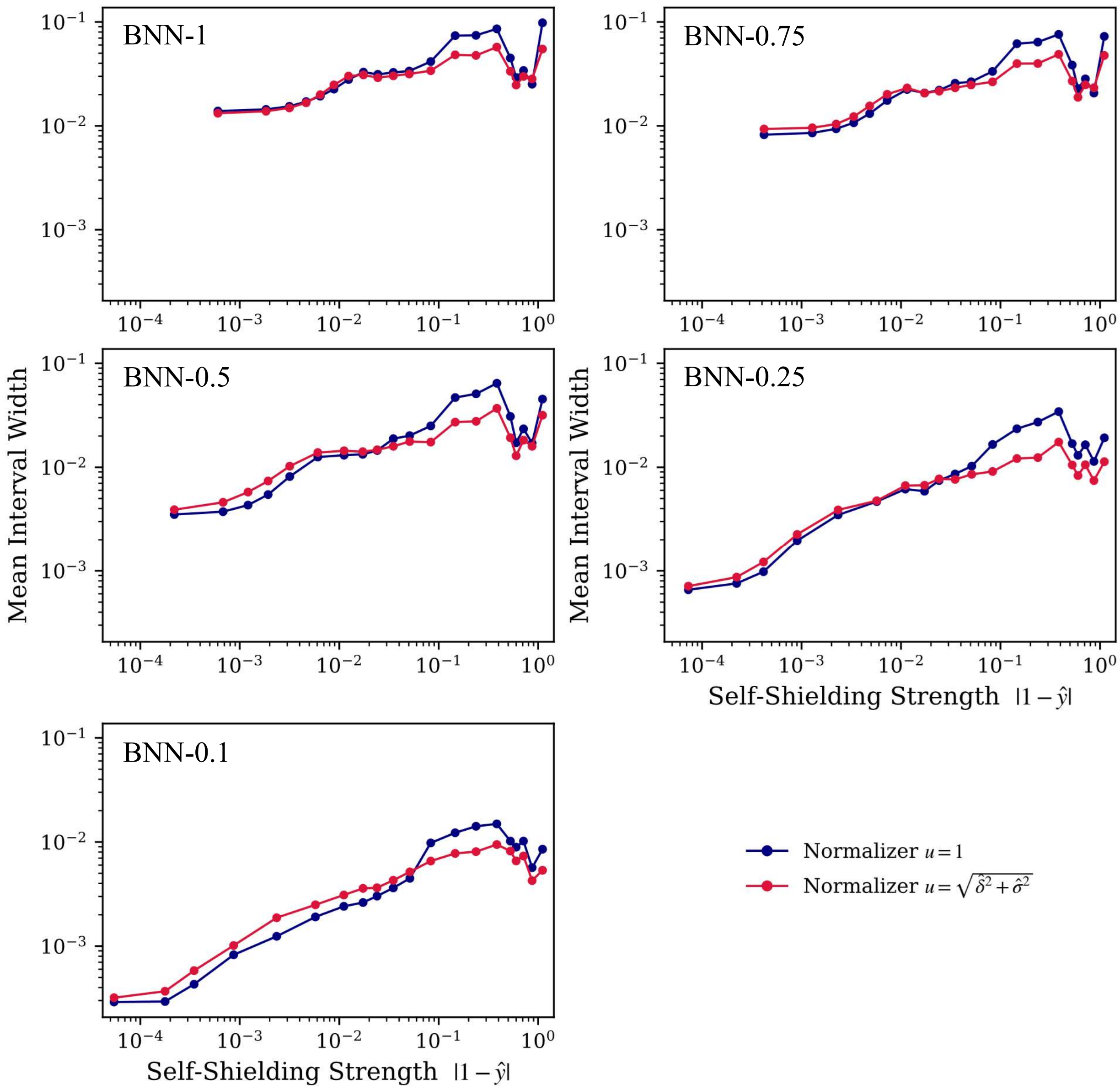

**Figure 16. Predicted interval widths for 95% target, CP stratified by self-shielding strength.** Individual plots correspond to BNNs with increasing levels of accuracy. The widths $w$ calculated using un-normalized CP, Eq. (7) and normalized CP, Eq. (10), are shown by the blue and red curves respectively. The self-shielding strength is greatest at the right-hand side of each plot, corresponding to small predicted cross-section ratios. Because the normalizer adjusts widths for individual samples, the widths shown in the normalized red curve represent the mean width in the self-shielding strength bin. The widths corresponding to the un-normalized blue curve are constant in each bin.

**Table V. Summary statistics, CP stratified by self-shielding strength (lower is better).**

| | **Mean width, 99% coverage target** | | **Mean width, 95% coverage target** | |
|---|---|---|---|---|
| Target MRE | Unnormalized | Normalized | Unnormalized | Normalized |
| 1.0% | $9.864\times10^{-2}$ | $4.137\times10^{-2}$ | $3.841\times10^{-2}$ | $3.069\times10^{-2}$ |
| 0.75% | $8.461\times10^{-2}$ | $3.409\times10^{-2}$ | $3.016\times10^{-2}$ | $2.435\times10^{-2}$ |
| 0.5% | $6.592\times10^{-2}$ | $2.490\times10^{-2}$ | $2.192\times10^{-2}$ | $1.649\times10^{-2}$ |
| 0.25% | $3.293\times10^{-2}$ | $1.180\times10^{-2}$ | $1.147\times10^{-2}$ | $7.512\times10^{-3}$ |
| 0.1% | $1.660\times10^{-2}$ | $8.232\times10^{-3}$ | $5.789\times10^{-3}$ | $4.500\times10^{-3}$ |

4.3.2 Stratification by output channel

Figures 17 and 18 show the coverage of CP stratified by output channel, as a function of the binned self-shielding strength for a target coverage of 99% and 95% respectively. The self-shielding strength was chosen as abscissa here because the output channel is a categorical variable, making visualization, and direct comparison with unstratified CP and stratification on self-shielding strength difficult. The unnormalized coverage at both 99% and 95% targets is more balanced than the coverage provided by unstratified, unnormalized CP, Figs. 9,10. The balance of the coverage at both target levels, however, is not significantly improved by the normalizer.

Figures 19 and 20 show the mean interval width for CP stratified by output channel. Again the results are shown as a function of the self-shielding strength for target coverage levels of 99% and 95% respectively. The inclusion of the normalizer provides a slight reduction in predicted intervals widths overall, with the effect being more pronounced for BNNs with higher MRE values. The results suggest that the output channel axis is responsible for most of the variation in the overall uncertainty in the BNNs' predictions. Results from Figs. 19 and 20 are summarized in Table VI.

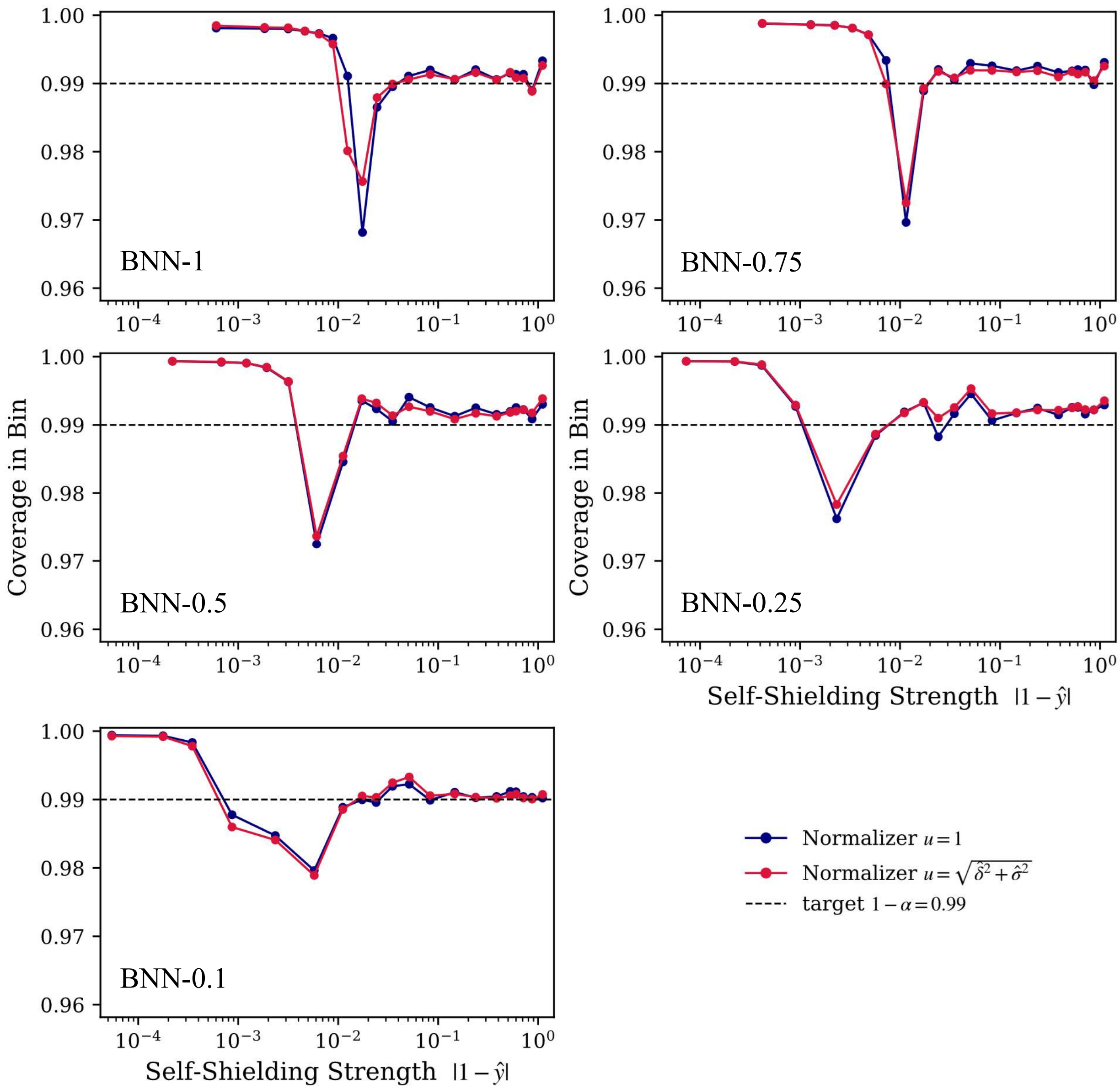

**Figure 17. Coverage level for 99% target, CP stratified by BNN output channel.** Individual plots correspond to BNNs with increasing levels of accuracy. The coverage calculated using un-normalized CP, Eq. (7) and normalized CP, Eq. (10), are shown by the blue and red curves respectively. The dashed line represents the target coverage level. The self-shielding strength is greatest at the right-hand side of each plot, corresponding to small predicted cross-section ratios.

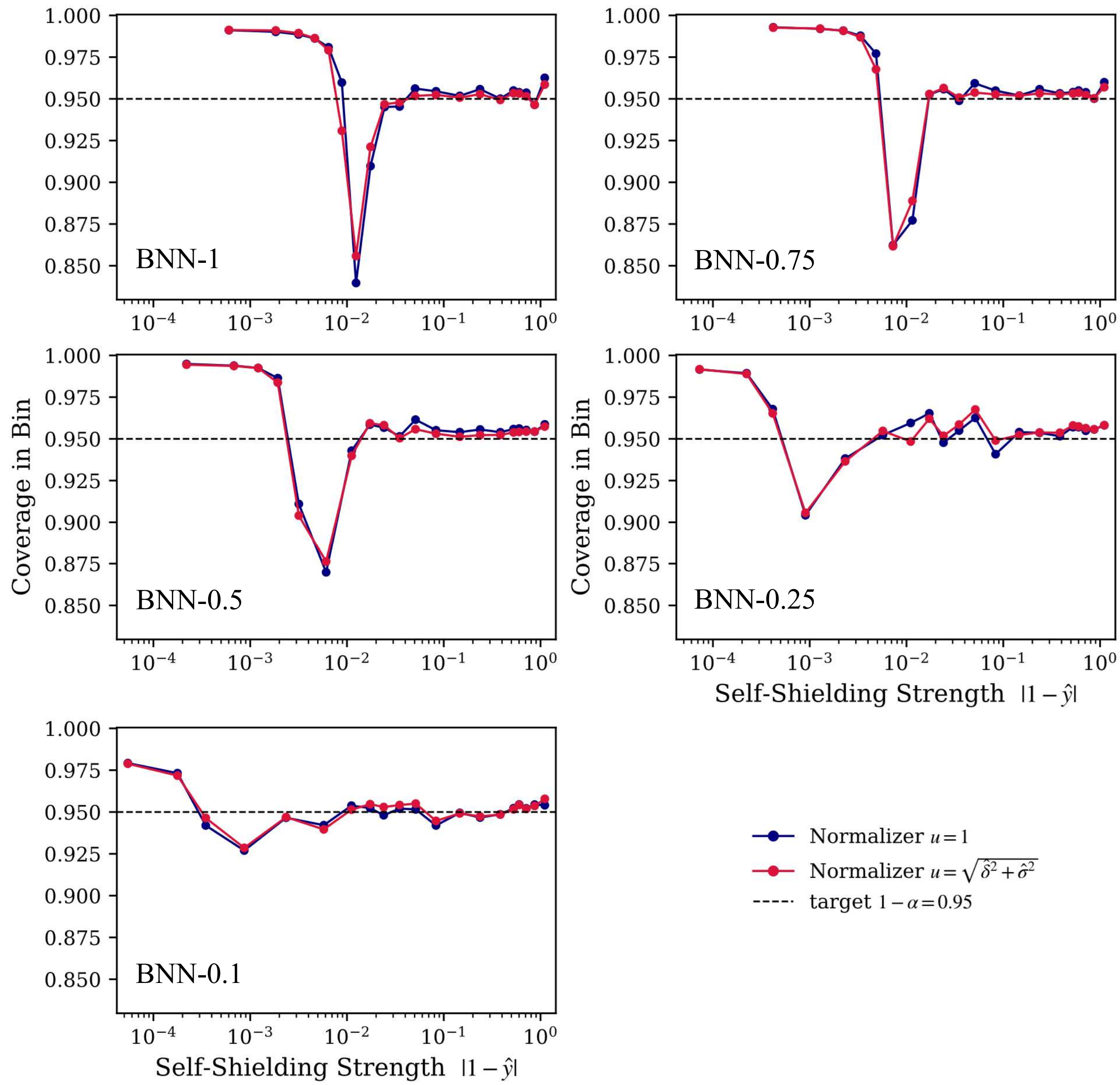

**Figure 18. Coverage level for 95% target, CP stratified by BNN output channel.** Individual plots correspond to BNNs with increasing levels of accuracy. The coverage calculated using un-normalized CP, Eq. (7) and normalized CP, Eq. (10), are shown by the blue and red curves respectively. The dashed line represents the target coverage level. The self-shielding strength is greatest at the right-hand side of each plot, corresponding to small predicted cross-section ratios.

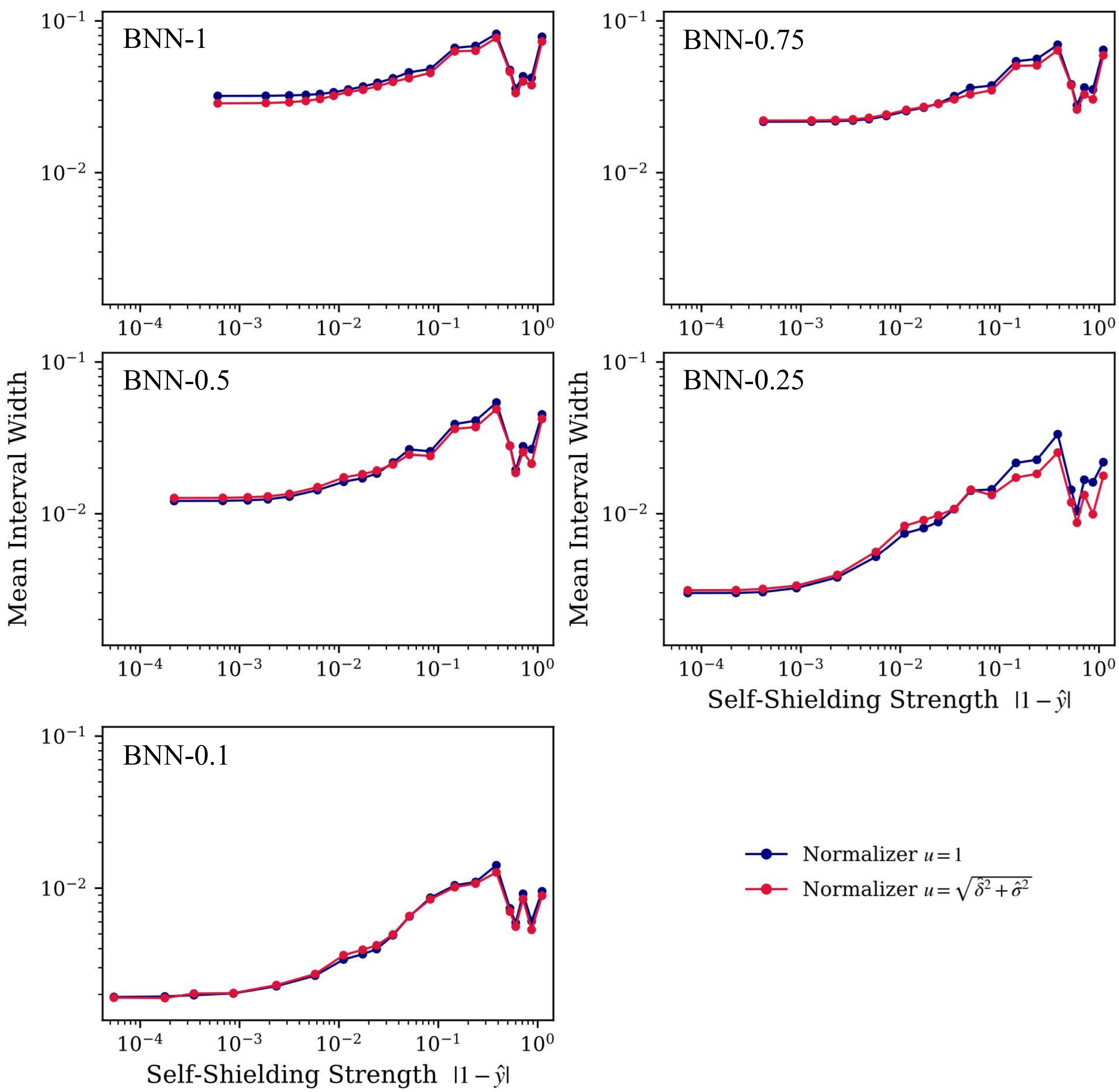

**Figure 19. Predicted interval widths for 99% target, CP stratified by BNN output channel.** Individual plots correspond to BNNs with increasing levels of accuracy. The widths $w$ calculated using un-normalized CP, Eq. (7) and normalized CP, Eq. (10), are shown by the blue and red curves respectively. The self-shielding strength is greatest at the right-hand side of each plot, corresponding to small predicted cross-section ratios. The widths corresponding to both curves shown represent the mean in each self-shielding strength bin.

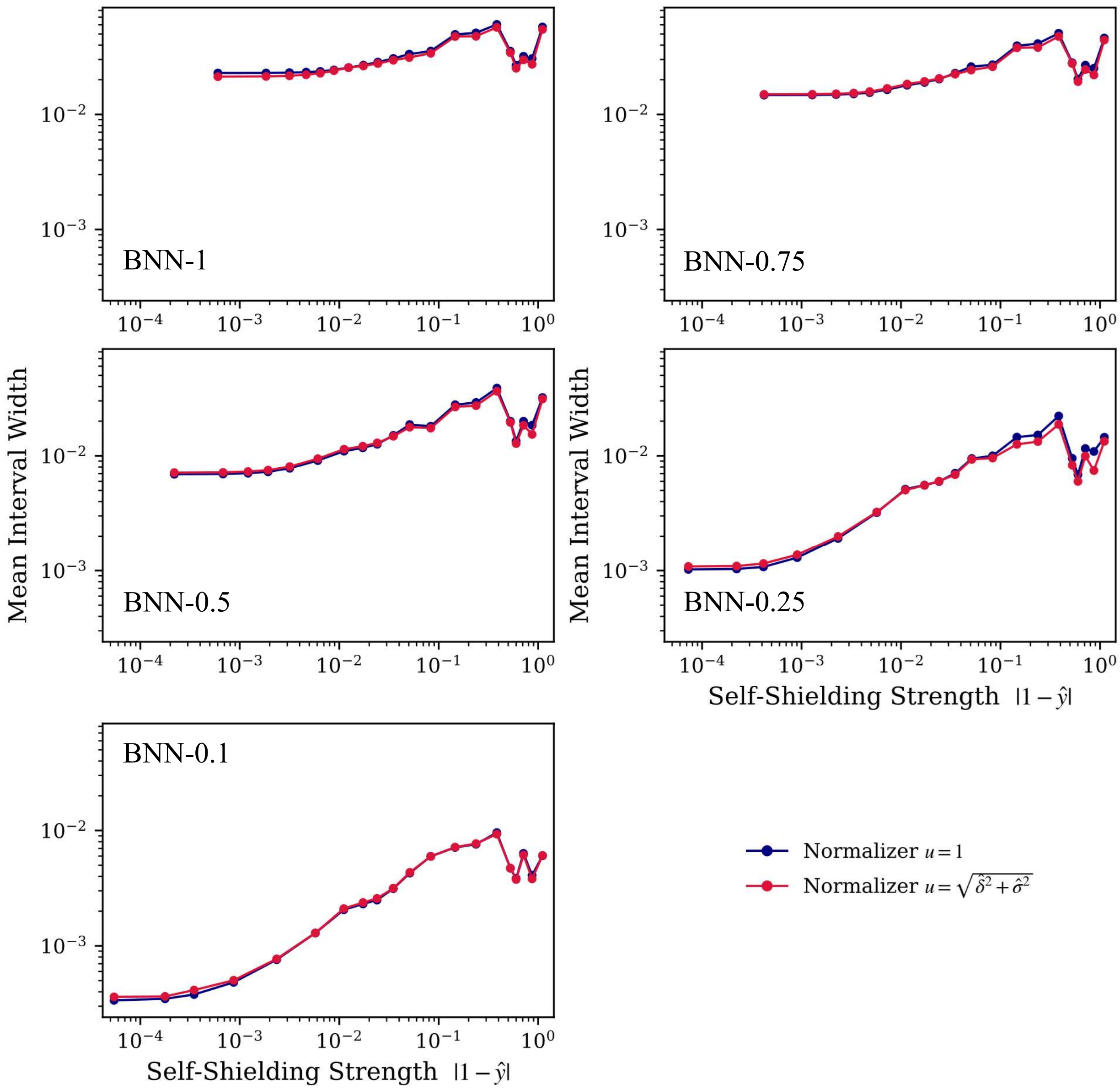

**Figure 20. Predicted interval widths for 95% target, CP stratified by BNN output channel.** Individual plots correspond to BNNs with increasing levels of accuracy. The widths $w$ calculated using un-normalized CP, Eq. (7) and normalized CP, Eq. (10), are shown by the blue and red curves respectively. The self-shielding strength is greatest at the right-hand side of each plot, corresponding to small predicted cross-section ratios. The widths corresponding to both curves shown represent the mean in each self-shielding strength bin.

**Table VI. Summary results, CP stratified by output channel (lower is better).**

| | Mean width, 99% coverage target | | Mean width, 95% coverage target | |
|---|---|---|---|---|
| Target MRE | Unnormalized | Normalized | Unnormalized | Normalized |
| 1.0% | $4.537\times10^{-2}$ | $4.236\times10^{-2}$ | $3.318\times10^{-2}$ | $3.155\times10^{-2}$ |
| 0.75% | $3.511\times10^{-2}$ | $3.337\times10^{-2}$ | $2.509\times10^{-2}$ | $2.426\times10^{-2}$ |
| 0.5% | $2.417\times10^{-2}$ | $2.311\times10^{-2}$ | $1.658\times10^{-2}$ | $1.602\times10^{-2}$ |
| 0.25% | $1.211\times10^{-2}$ | $1.052\times10^{-2}$ | $7.895\times10^{-3}$ | $7.094\times10^{-3}$ |
| 0.1% | $5.879\times10^{-3}$ | $5.677\times10^{-3}$ | $3.659\times10^{-3}$ | $3.644\times10^{-3}$ |

4.4 Stratified Conformal Prediction: Overshoot Severity

Because CP only guarantees coverage to a specific target, a minority of predicted self-shielding factors will inevitably fall outside predicted interval widths. The severity of these overshoots is defined according to Eq. (21):

$$\rho = \frac{|\hat{y} - y|}{w_{1/2}} \quad (21).$$

where $w_{1/2}$ is the prediction interval half-width. In this section we compare the overshoot severity using un-normalized, i.e. $u_i = 1$, and normalized, i.e. Eq. (20) CP. Figures 21 and 22 show the fraction of overshoots in the predicted self-shielding factors as a function of overshoot severity, Eq. (21), for target coverage levels of 99% and 95% respectively. Here the results shown correspond to CP stratified along the predicted self-shielding strength. Equivalent results for CP stratified along the BNN output channel are shown in Figs. 23 (99% target coverage) and 24 (95% target coverage). The maximum overshoot values of Figs. 21-24 are given in Tables VII and VIII. The results show that normalizing using Eq. (20) reduces the severity of overshoots in almost all cases, with few exceptions. The reduction is most significant in CP stratified by self-shielding strength, at lower BNN accuracy levels. In CP stratified by output channel, the reduction in severity is most significant at higher BNN accuracy levels.

Figures 21-24 also show that extreme overshoots are rare but become more common as the BNN accuracy increases. This is not unexpected as predicted widths are narrower in more accurate BNNs. Additionally, as the BNNs' model error becomes smaller, the remaining overall uncertainty in predictions becomes dominated by the uncertainty in the ground truth data. Because of this, in the MC error dominated region there are more samples in the data whose model uncertainty can't be predicted reliably by the $\delta$-UNN. Consequently, the normalization factor $u_i$, Eq. (20), becomes less informative. There are also significantly fewer training examples for the UNNs where the residuals of BNN-0.1 are larger than the MC uncertainties, which reduces the UNNs' predictive ability. These combined effects are the likely causes of the normalizer's inability to reduce the severity of the largest extreme overshoots.

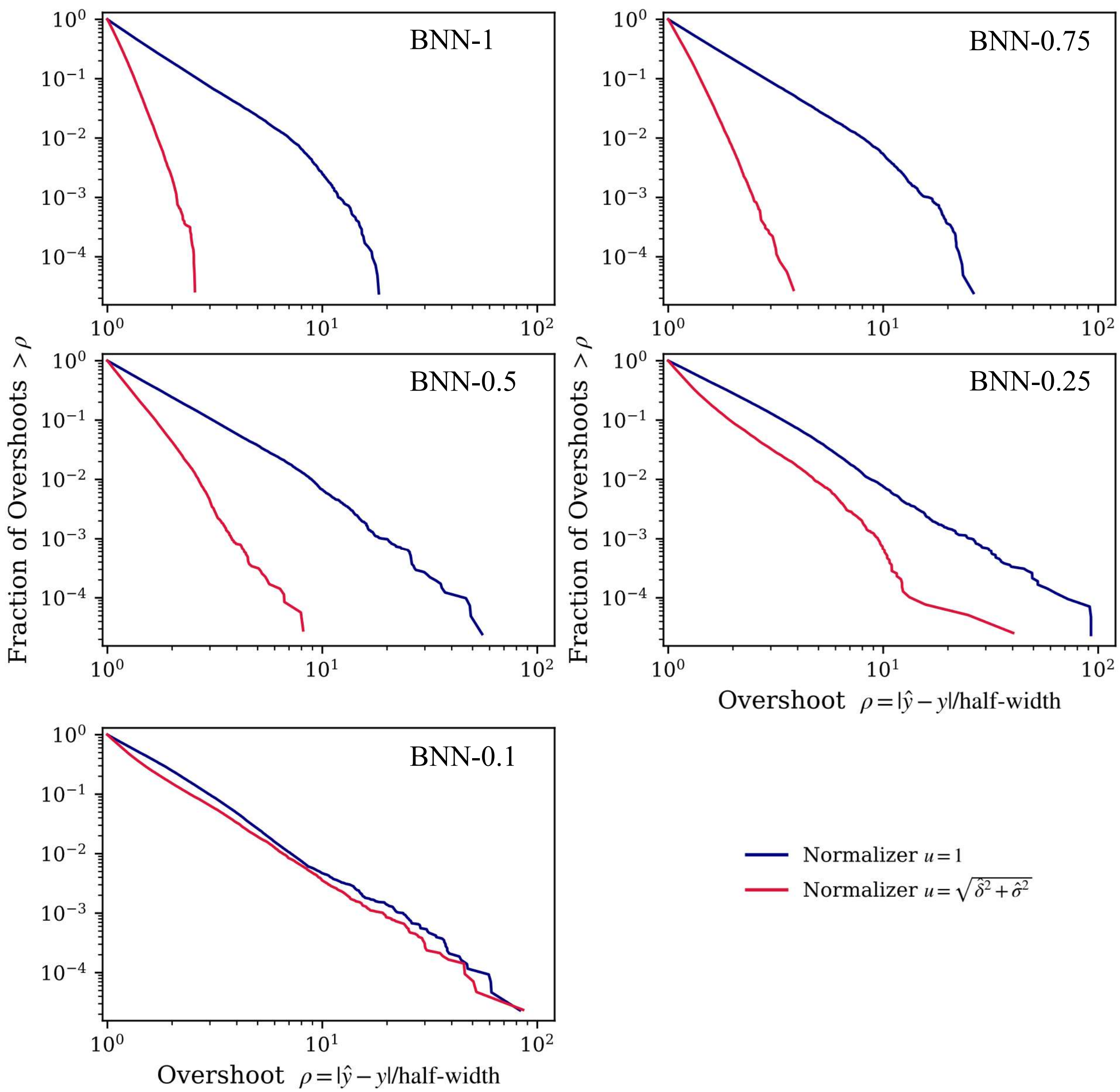

**Figure 21. Fraction of overshot self-shielding predictions as a function of the overshoot severity.** Results shown for stratified CP according to self-shielding strength, at 99% target coverage level. Individual plots correspond to BNNs with increasing levels of accuracy. The un-normalized and normalized CP results are shown by the blue and red curves respectively.

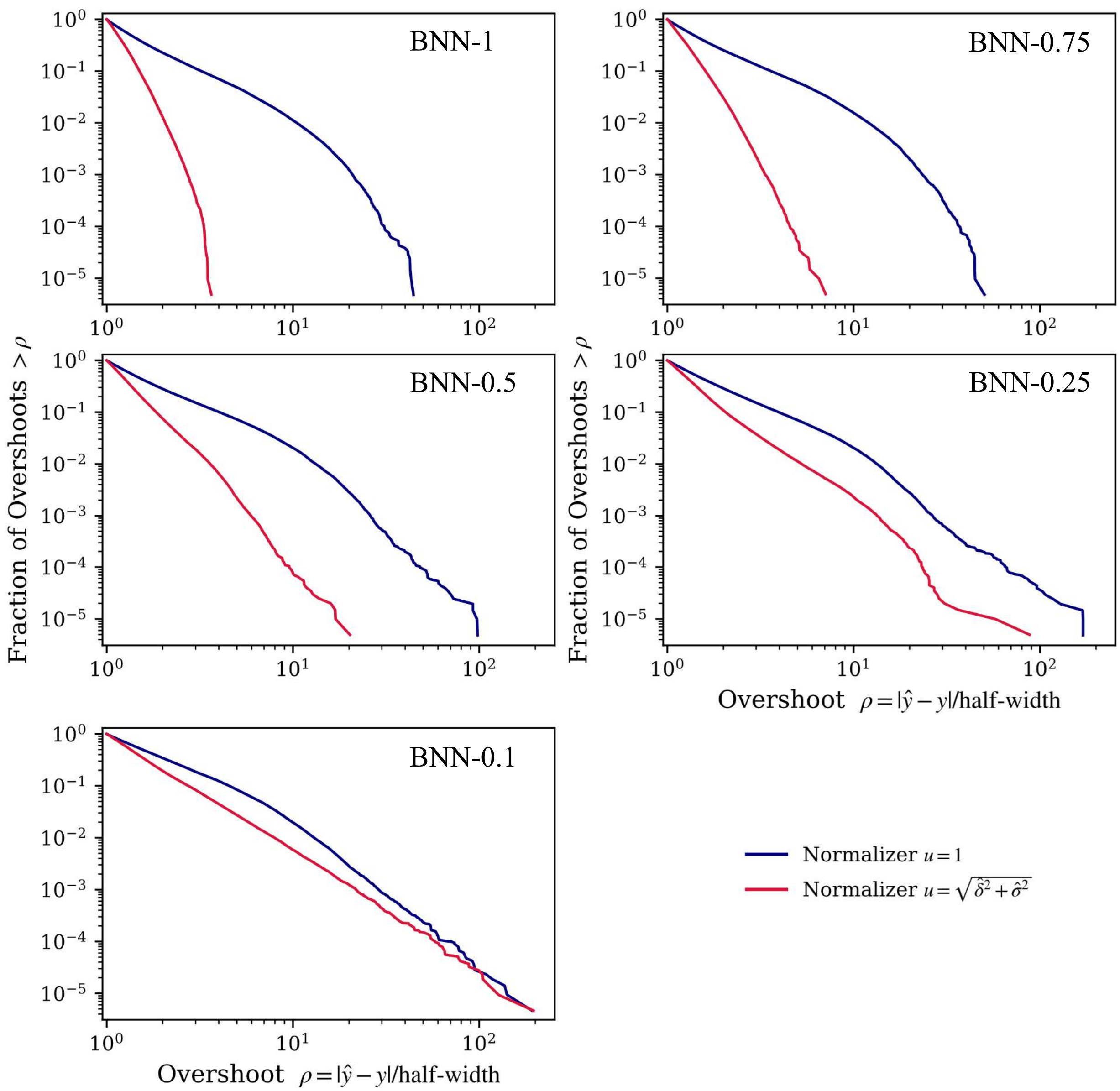

**Figure 22. Fraction of overshot self-shielding predictions as a function of the overshoot severity.** Results shown for stratified CP according to self-shielding strength, at 95% target coverage level. Individual plots correspond to BNNs with increasing levels of accuracy. The un-normalized and normalized CP results are shown by the blue and red curves respectively.

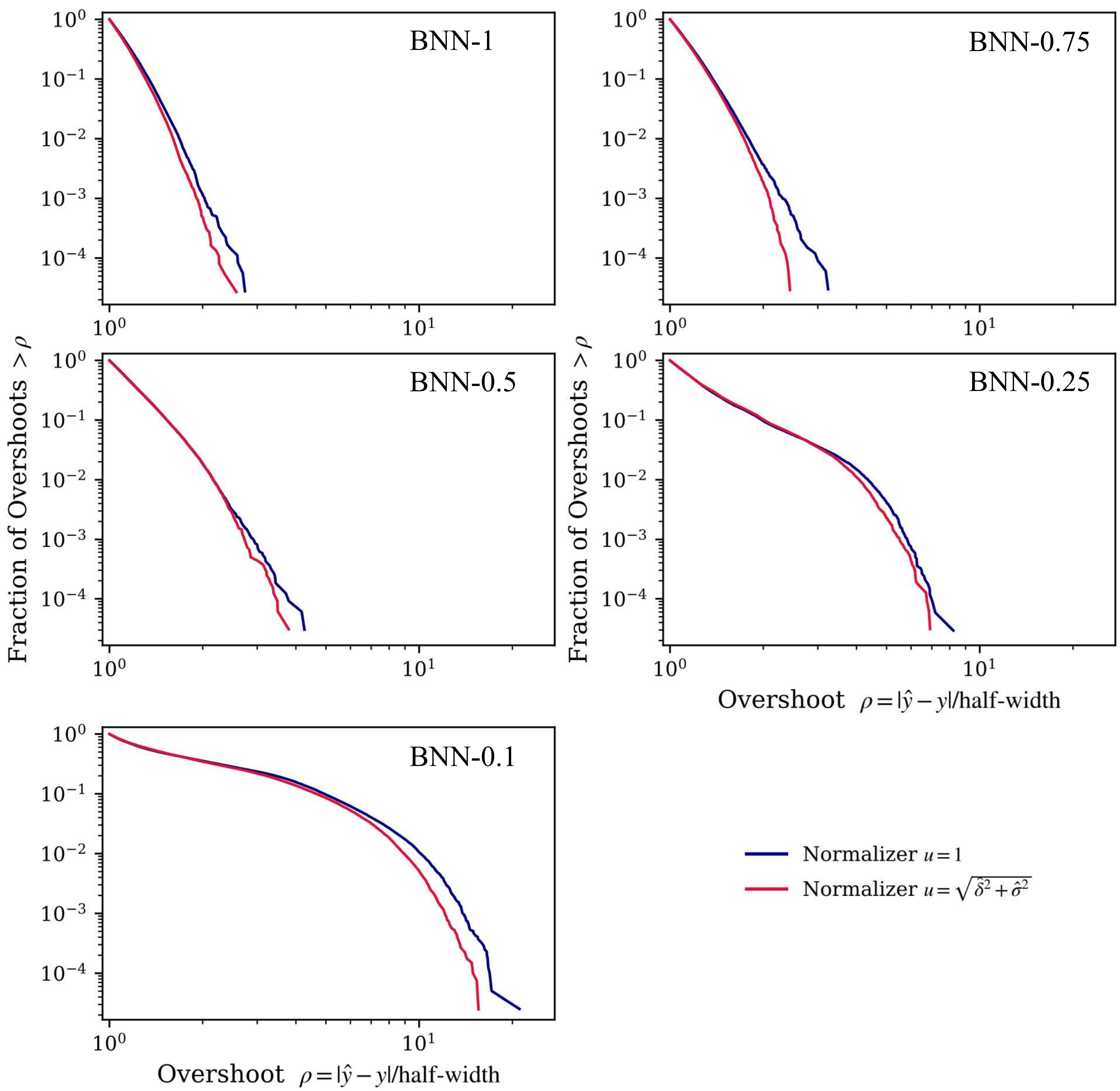

**Figure 23. Fraction of overshot self-shielding predictions as a function of the overshoot severity.** Results shown for stratified CP according to BNN output channel, at 99% target coverage level. Individual plots correspond to BNNs with increasing levels of accuracy. The un-normalized and normalized CP results are shown by the blue and red curves respectively.

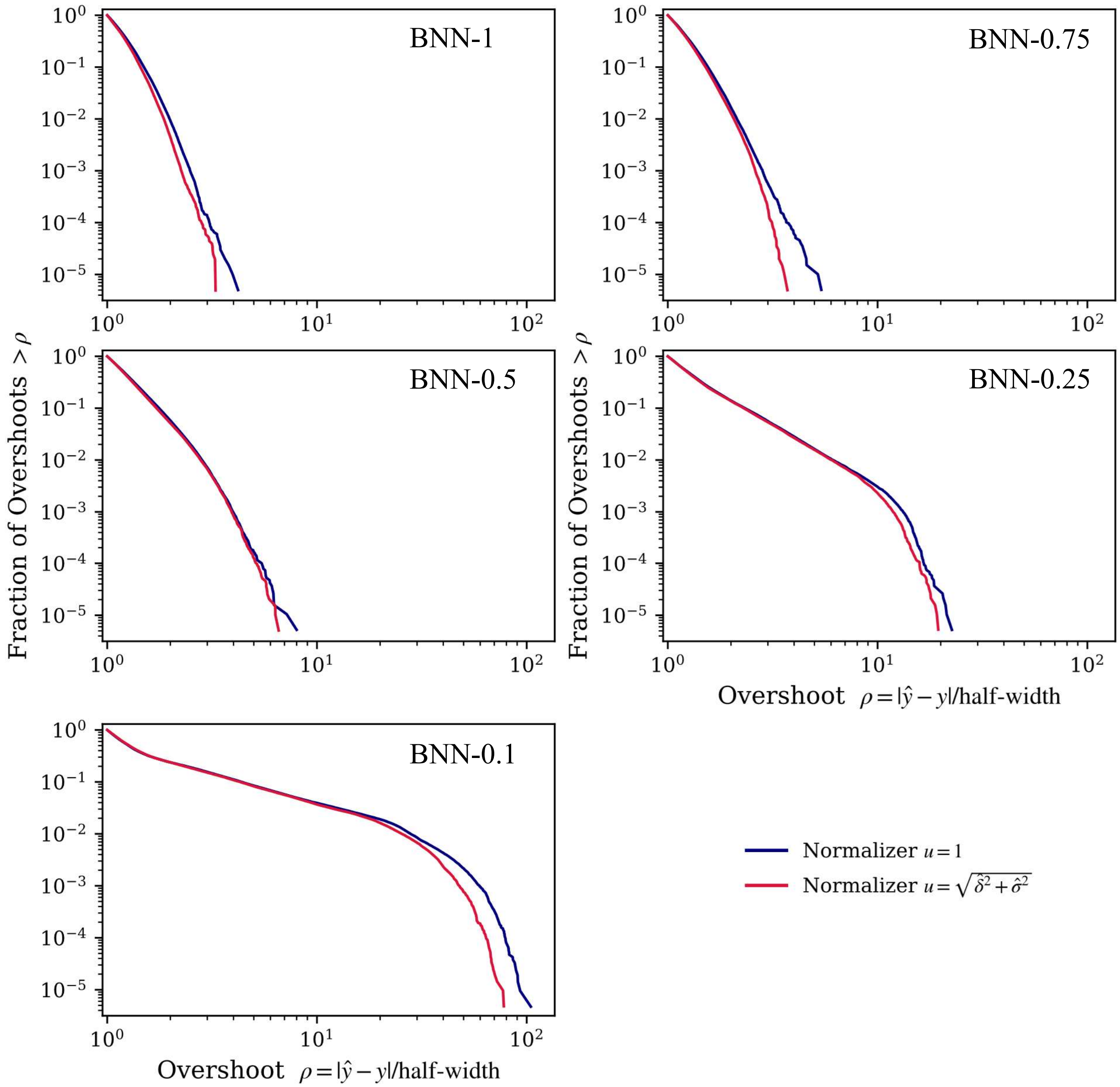

**Figure 24. Fraction of overshot self-shielding predictions as a function of the overshoot severity.** Results shown for stratified CP according to BNN output channel, at 95% target coverage level. Individual plots correspond to BNNs with increasing levels of accuracy. The un-normalized and normalized CP results are shown by the blue and red curves respectively.

**Table VII. Summary results, overshoot analysis, CP stratified on self-shielding strength (lower is better).**

| Target MRE | Maximum Overshoot $\rho$, 99% Target Coverage | | Maximum Overshoot $\rho$, 95% Target Coverage | |
|---|---|---|---|---|
| | Unnormalized (Normalized same-sample) [a] | Normalized (Unnormalized same-sample) [b] | Unnormalized (Normalized same-sample) [a] | Normalized (Unnormalized same-sample) [b] |
| 1.0% | 18.3 (1.2) | 2.6 (1.47) | 44.3 (1.3) | 3.6 (2.4) |
| 0.75% | 26.3 (1.0) | 3.8 (2.2) | 50.4 (1.8) | 7.1 (4.4) |
| 0.5% | 55.4 (2.2) | 8.1 (3.8) | 97.9 (4.0) | 20.2 (5.6) |
| 0.25% | 92.6 (10.4) | 40.2 (13.9) | 170.6 (23.1) | 88.0 (24.8) |
| 0.1% | 83.0 (19.8) | 85.9 (28.0) | 191.0 (45.1) | 195.5 (64.3) |

a) The normalized same-sample quantity in parentheses is the overshoot when using the normalizer, for the same test sample in both normalized and unnormalized cases. This is provided to indicate that the table does not show a direct reduction between un-normalized and normalized CP on a sample-wise basis, only the worst results from each set.
b) The unnormalized same-sample quantity in parentheses is the converse of the previous footnote.

**Table VIII. Summary results, overshoot analysis, CP stratified on output channel (lower is better).**

| Target MRE | Maximum Overshoot $\rho$, 99% Target Coverage | | Maximum Overshoot $\rho$, 95% Target Coverage | |
|---|---|---|---|---|
| | Unnormalized (Normalized same-sample) [a] | Normalized (Unnormalized same-sample) [b] | Unnormalized (Normalized same-sample) [a] | Normalized (Unnormalized same-sample) [b] |
| 1.0% | 2.7 (1.0) | 2.6 (1.9) | 4.2 (1.4) | 3.3 (2.1) |
| 0.75% | 3.2 (0.8) | 2.4 (2.1) | 5.4 (1.1) | 3.7 (3.6) |
| 0.5% | 4.3 (0.6) | 3.8 (2.3) | 8.0 (1.0) | 6.6 (5.7) |
| 0.25% | 8.2 (6.2) | 6.9 (4.8) | 22.6 (19.5) [c] | 19.5 (22.6) [c] |
| 0.1% | 21.1 (15.6) [c] | 15.6 (21.1) [c] | 104.6 (77.8) [c] | 77.6 (104.8) [c] |

a) The normalized same-sample quantity in parentheses is the overshoot when using the normalizer, for the same test sample in both normalized and unnormalized cases. This is provided to indicate that the table does not show a direct reduction between un-normalized and normalized CP on a sample-wise basis, only the worst results from each set.
b) The unnormalized same-sample quantity in parentheses is the converse of the previous footnote.
c) Here the normalized and unnormalized maximum overshoot correspond to the same sample.

## 5. Conclusions

UNNs were trained to predict model and MC uncertainties using the penultimate layer activations of pre-trained BNNs as inputs. The BNNs were trained to predict self-shielding factors in microscopic multigroup cross sections with target accuracies from 1% to 0.1% MRE. The model and MC uncertainties emitted by the UNNs were combined in quadrature to provide a normalization factor for split CP and stratified CP. In the latter case the stratification was done along the self-shielding strength axis and the output channel. We compared the balance in coverage and the predicted interval widths in unnormalized and normalized CP. In the stratified cases we also compared the severity of overshoots with and without the normalizer.

Results from our reliability analysis and standardized residuals show that $\delta$-UNNs make predictions of model error that are well-calibrated in the region where the model uncertainty dominates. In the MC uncertainty dominated region, however the standardized residuals indicate that the $\delta$-UNNs are unable to reliably predict model error in the predictions of BNNs with low MRE values. This is due to the model error being unresolvable when the MC error is large, and BNN residuals are smaller than the combined model and MC error. The results also reflect the limitation in the models' predictive ability when ground truth uncertainties are propagated assuming reaction rate and flux tally uncertainties are independent. The predicted MC uncertainties, while well-calibrated, show a relatively large spread, which is also likely due to inaccurate ground truth MC uncertainties in the training data.

Despite these limitations, CP is dramatically improved overall when the predicted uncertainties are combined in quadrature and used as a CP normalizer. The effect of the normalizer on unstratified CP improves balance in coverage and reduces interval widths significantly, particularly for higher MRE BNNs at 99% target coverage. The improvement is modest at 95% target coverage and for the more accurate BNNs. This pattern in reduction in interval widths remains consistent in stratified CP, particularly when stratification is along self-shielding strength. In CP stratified along the output channel, the normalizer reduces interval widths modestly but consistently. In almost all cases, the normalizer reduces the severity of BNN predictions overshooting the predicted widths. In contrast to the effect the normalizer has on widths, the reduction in overshoot severity is consistently better at low BNN MRE values when CP is stratified by output channel.

This work suggests that the proposed approach to normalization in CP is robust across a useful range of ANN based point predictor accuracy levels. Because the method can be combined with existing point predicting models, evaluating its robustness in different contexts should not require significant effort. Because the normalizer's benefit was only modest at low BNN MRE levels, the focus of follow-up studies will be to investigate ways to improve the normalizer for accurate BNNs. A clear path towards this is to re-train the BNNs using accurate ground truth uncertainties before training the UNNs. OpenMC simulations, however, will need to be repeated to access the correlation factors that will enable correct tally uncertainty propagation. Future studies should also prioritize investigating the effect of the normalizer under covariate shift. A benefit of CP is that coverage remains guaranteed as long as new data points originate from the same distribution as the calibration data. It is unclear whether the normalizer will retain its utility if new data points are out-of-distribution with respect to the data on which the UNNs were trained.

**Acknowledgments**

This work was supported by DOE-NE award DE-NE0009418.

**Appendix A. Empirical Model Uncertainties.**

The true model error $\delta(x_i)$ is the difference in the model prediction $\hat{y}(x_i)$ and the true value $\mu(x_i)$:

$$\delta xi{=}yxi{-}\mu xi \quad \text{(A1).}$$

Because the true value is inaccessible, however, the empirical model error can be estimated using the Monte Carlo ground truth values. We assume the Monte Carlo ground truth values $y(x_i)$ can be decomposed in terms of the true value $\mu(x_i)$ and Monte Carlo noise $n(x_i)$:

$$y(x_i) = \mu(x_i) + n(x_i) \quad \text{(A2).}$$

where:

$$n(x_i) \sim \mathcal{N}(0, [\sigma_{omc}(x_i)]^2) \quad \text{(A3).}$$

Using Eq. (A2) we can write the residual:

$$r(x_i) = \hat{y}(x_i) - y(x_i) = [\hat{y}(x_i) - \ \mu(x_i)] - n(x_i) \quad \text{(A4).}$$

It can be shown that squaring Eq. (A4) and taking the expectation value gives Eq. (13), which can be rearranged to solve for the variance associated with the model error, $\delta(x_i)^2$:

$$\delta(x_i)^2 = \mathbb{E}[r^2 \mid x_i] - [\sigma_{omc}(x_i)]^2 \quad \text{(A5).}$$

Because the expectation from a single draw $i$, however, is unobservable, it is estimated by averaging over many samples. Samples are selected by binning according to similarity in their predicted model error $\hat{\delta}$. For a bin $B$ with $N_B$ samples:

$$\delta_B^2 = max\left(0, \frac{1}{N_B}\sum_{i\in B} r^2(x_i) - \frac{1}{N_B}\sum_{i\in B}[\sigma_{omc}(x_i)]^2\right) \quad \text{(A6).}$$

which can be written as Eq. (16), where the maximum is taken to account for residuals smaller than the OpenMC uncertainty.

**CRediT authorship contribution statement**

**Joshua Nichols:** Data Curation, Software, Writing – Review & Editing. **Andrew Osborne**: Conceptualization, Methodology, Software, Writing – Original Draft. **Valerio Mascolino:** Writing – Review & Editing.

**Data availability**

Computer code and data used in this work are available upon reasonable request.

**Declaration of generative AI and AI-assisted technologies in the manuscript preparation process**

During the preparation of this work, the authors used Anthropic's Claude Opus 4.8 and earlier model versions to assist in the generation of visualization code, the analysis code and ML model implementations in PyTorch. Opus 4.8 was also used to assist in literature searches, mathematical formulation and preliminary error checking. After using this tool/service, the authors reviewed and edited the content as needed and take full responsibility for the content of the published article.